\documentclass{aa}
\usepackage{natbib}
\usepackage{amssymb,verbatim}
\usepackage{amsmath}
\usepackage{graphicx}
\usepackage{appendix}
\usepackage{subfig}
\usepackage{color}

\begin{document}
\title{The evolving cluster cores: Putting together the pieces of the puzzle}


\author{S. Molendi\inst{1},  S. De Grandi\inst{2}, M. Rossetti\inst{1}, I. Bartalucci\inst{1}, F. Gastaldello\inst{1}, S. Ghizzardi\inst{1} and M.Gaspari\inst{3} }

\offprints{S. Molendi \email{silvano.molendi@inaf.it}}

\institute{INAF - IASF Milano, via A.Corti 12 I-20133 Milano, Italy \and
INAF - Osservatorio Astronomico di Brera, via E.Bianchi 46, 23807 Merate, Italy \and
Department of Astrophysical Sciences, Princeton University, Princeton, NJ 08544, USA 
}
\date{\today}
\abstract
{In this work we address the issue of whether the division of clusters in cool cores (CCs) and non-cool cores (NCCs) is due to a primordial difference or to how clusters evolve across cosmic time.}
{Our first goal is to establish if spectra from the central regions of a subclass of NCCs known as cool core remnants (CCRs) are consistent with having a small but significant amount of short cooling time gas, thereby allowing a transformation to CC systems on a timescale of a giga year. Our second goal is to determine if low ionization Fe lines emitted from this residual cool gas will be detectable by the calorimeters that will fly on board XRISM and ATHENA.}
{We performed a spectral analysis of CCR systems with a multi temperature model and, assuming the different components to be in pressure equilibrium with one another, derived entropy and cooling time distributions for the X-ray emitting gas.} 
{We find that in most of our systems, the spectral model allows for a fraction of low entropy, short cooling time gas with a mass that is comparable to the one in CC systems. Moreover, simulations show that future spectrometers on board XRISM and ATHENA will have the power to directly resolve emission lines from the low temperature gas, thereby providing incontrovertible evidence for its presence.}
{Within the scenario that we have explored, the constant fraction of CCs measured across cosmic time emerges from a dynamical equilibrium where CCs transformed in NCCs through mergers are balanced by NCCs that revert to CCs. Furthermore, CCs and NCCs should not be viewed as distinct sub classes, but as ``states" between which clusters can move. }
\keywords{galaxies: clusters: intracluster medium -- X-ray: galaxies: clusters -- intergalactic medium}

\titlerunning{Evolving Cluster Cores}

\authorrunning{Molendi et al.}

\maketitle

\section{Introduction}\label{sec:intro}

Despite the impressive progress made over the last several decades,  our understanding of the physics of the intra-cluster medium (ICM) is, in many  ways, limited.  Just to mention one example that is particularly relevant to this work, we do have evidence that thermal conduction is significantly suppressed (see Sect.\ref{sec:phys} for a discussion), but we have limited knowledge of the physical processes that are responsible for this phenomenon. Similarly we have proof that the ICM is turbulent \citep[e.g.,][]{Khatri:2016,Eckert:2017b}, but we do not know on what scales and by what mechanism the turbulence is dissipated. 
One consequence of these constraints is that they limit  our ability to turn observational classifications of clusters into  physically motivated categories.

In this paper we consider one particular way of classifying galaxy clusters, namely if they are cool cores (CCs) or non-cool cores (NCCs). A CC cluster \citep{Molendi:2001} typically
displays a central surface brightness enhancement  accompanied by a reduction in temperature \citep[e.g.,][]{Leccardi_temp:2008,Cavagnolo:2009,Arnaud:2010,Mcdonald:2013}  and an increase in metallicity \citep{DGM:2001,Leccardi:2010}. Over the last decade, there has been considerable discussion on how exactly on should assign objects to the two classes, use of central cooling time or entropy \citep[e.g.,][]{Hudson:2010,Mcdonald:2013} are among the most popular solutions. Another point that has been debated is if the  two classes should be treated as distinct populations or the extremes of a unique distribution \citep[e.g.,][]{Cavagnolo:2009,Pratt:2010}.
The lack of clear-cut separation between CC and NCC properties suggests that they fall into the second category.
 Moreover, while objects with marked CC or NCC  properties are correctly classified by any of these definitions, intermediate objects may end up being classified differently by different authors. 
 
From the theory and simulation side there has been an effort from several authors to understand if mergers can disrupt CCs, thereby turning CCs into NCC systems \citep[e.g.,][]{McCarthy:2008,Rasia:2015,Barnes:2018}. However, since key processes such as conduction and mixing operate on physical scales that are orders of magnitude smaller than those that can be sampled by simulations, they are introduced in the form of sub grid physics through simple recipes which may well fail to provide an adequate description.
In light of these limitations on the simulation side, an observational approach is all the more important.
As we shall see further on, the limited understanding of physical processes in the ICM  plays an important role in the relationship between CC and NCC systems. For instance, diverse active galactic nuclei, AGN, feedback/feeding models shape differently the CC-NCC distributions and related properties such as the cooling time  \citep[e.g.][and refs within]{Gaspari:2020}.

The structure of this paper is somewhat unusual. Typically, observational manuscripts first present data and then discuss implications while theory or simulation works first present a model and then compare predictions with observations.
Here we start off by presenting available observational constraints on the relationship between CC and NCC systems.
We then discuss an apparent contradiction between observational properties and propose a solution requiring a change in the description of ICM properties in cluster cores. From this we build a spectral model which we compare with available observational data.   In the hope of further clarifying our approach we provide a more detailed description. In Sect.\ref{sec:cc_ncc} we discuss the relationship between CCs and NCCs. We show that while we have several indications that CCs can been turned into NCCs systems, there is conflicting evidence regarding the opposite process that is to say NCCs changing into CCs.  
In Sect.\ref{sec:phys} we discuss how inhibition of thermal conduction may resolve the conflict. Indeed, if the ICM in NCC cores is characterized by a distribution of temperatures poorly approximated by a single one,  the standard estimator of the time required to turn a NCC into a  CC, i.e. the cooling time, will be miscomputed.
In Sect.\ref{sec:con_spec} we construct a spectral model that allows for a distribution of temperatures and apply it to
a sample of NCC spectra. We find that, in most cases, the spectral model allows for a fraction of cool, low entropy and short cooling time gas
with a mass that is  comparable to the one found in CC systems. In Sect.\ref{sec:discussion} we discuss our results and  in Sect.\ref{sec:future} we employ  simulations
to show how future spectrometers on board XRISM and ATHENA will have the power to directly resolve emission lines from the low temperature gas thereby providing incontrovertible evidence for its presence. Finally in Sect.\ref{sec:summary} we summarize our main findings.
Throughout the paper we assume a $\Lambda$ cold dark matter cosmology with $H_o = $70 km s$^{-1}$ Mpc$^{-1}$, $\Omega_M = 0.3$ and $\Omega_\Lambda = 0.7$.

\section{Understanding the CC/NCC division}\label{sec:cc_ncc}

Several lines of evidence have emerged over the last decade suggesting that: 1) the environment in which CC and NCC form and  live are indistinguishable; 2) CC systems are disrupted by mergers. We briefly review them in this section.

\subsection{Evidence for CC disruption}
If CCs have always been CCs they should somehow be different from other clusters not just in the core but also further out, however we have found no evidence of this. Studies of clusters on large scales show no obvious difference between CCs and NCCs. \citet{Ghirardini:2019,McDonald:2017} find no difference between CCs and NCCs in terms of the  radial profiles of thermodynamic quantities at large radii. A similar result has been found for metal abundance profiles, see \citet{Ghizzardi:2021}. Moreover, in a recent study \citet{Medezinski:2017,Medezinski:2019}, measuring the assembly bias using both clustering and weak lensing, find no difference between CC and NCC systems. All these results disfavor ``ab initio" models. 

It should be pointed out that different authors use different criteria to divide CC from NCC systems: \cite{Ghirardini:2019} used the central entropy value  (and a threshold value of 30 keV cm$^2$) as measured by Chandra and reported by \cite{Cavagnolo:2009}; \cite{McDonald:2017}  use the central density, avoiding the weak cool core regime as defined by \cite{Hudson:2010} and therefore applying the NCC recipe for 
$n_{e,0} < 0.5 \cdot 10^{-2}$ cm$^{-3}$ and CC for $n_{e,0} > 1.5  \cdot  10^{-2}$  cm$^{-3}$; \cite{Medezinski:2019} use the presence of strong $H_{\alpha}$ nebular luminosity. These definitions lead to consistent classifications in the case of  objects with marked CC or NCC traits, conversely intermediate or transitional objects can sometime end up being classified differently by different authors. 
		
There is a consistent body of observational results that, over the last 20 years, have connected giant radio halos (hereafter RHs) to recent mergers \citep[e.g.][]{Buote:2001,Brunetti:2009,Cassano:2010,Cassano:2013,Cuciti:2015}. 
What is found in all these works is that cluster wide RH are associated with dynamically disturbed systems. This observational result is at the basis of the widely accepted scenario of turbulent reacceleration for radio halo formation \citep{Brunetti:2014}.
In a paper from a few years ago \citep{Rossetti:2011}, found that in a well defined 
sample of X-ray selected clusters,  none of the objects hosting a RH can be classified as a CC, similar results were presented by \cite{Cassano:2010}, \cite{Cassano:2013} and \cite{Cuciti:2015}.
This suggests that the main mechanism that can start a large-scale synchrotron emission,  major mergers, is the same that can destroy CCs. 
 Several theory papers have been dealing with this issue, despite the considerable challenge of the complicated (sub grid) physics related to core balance and evolution/destruction which is still very difficult to capture in simulations. Initial works \citep{McCarthy:2004,McCarthy:2008,Burns:2008} suggested the CC and NCC systems followed distinct evolutionary tracks.
A long road of improvement traced by more recent results shows the effectiveness of mergers in transforming CCs into NCC both in a cosmological
setting \citep[e.g.,][and refs therein]{LeBrun:2014,Rasia:2015}  and in tailored simulations \citep[e.g.,][and refs therein]{ZuHone:2011,Valdarnini:2021}. Moreover, AGN feedback via outflows/jets is expected to further contribute to the CC-NCC transition via inside-out heating  \citep[e.g.]{Gaspari:2014b}.

In recent years, a few cases of CC clusters hosting Radio Halos have been reported \citep[e.g.]{Bonafede:2014,Savini:2018}, however upon closer inspection these objects, for one reason or another, fit with some difficulty either in the CC or in the RH category.  For example, in CC objects such as RX$\,$J1720.1$+$2638 and A2390 extended emission is not detected at 1.4 GHz 
but at lower frequencies. Moreover, in the former system, it has been associated to a Giant Radio Bubble or a minor 
merger \citep{Biava:2021} and in the latter, after a tentative RH classification \citep{Sommer:2017}, it has been attributed to a Giant Radio Galaxy \citep{Savini:2019}.

Furthermore, even if a small fraction of CC clusters were to host Radio Halos, as could be the case for CLJ1821+624 \citep{Bonafede:2014}, it would leave the general picture unchanged, i.e. the vast majority of CCs do not host RHs and mergers that generate RHs typically destroy CCs.
In \cite{Leccardi:2010}, we showed that, for a sample defined starting from the XMM-Newton archive, CCs are not found in systems that feature manifestations of a major merger, at radio, optical or X-ray wavelengths, again suggesting that mergers destroy CCs. We did find two exceptions, namely A115N and A85 both CCs and very disturbed systems. However, in both cases the evidence found in the literature supported a scenario where the effects of the merger have not reached the
core of the main structure, see  \citet{Leccardi:2010} for details. We consider this as a  warning: inferences drawn from the way objects are distributed into different classes need to be gauged within the astrophysical framework that is being tested, failing to do so can lead astray. 

Another important piece of the puzzle comes from \citet{Rossetti:2010}, hereafter RM10. In that paper we found that, for a well defined X-ray sample,  most NCC clusters  host regions reminiscent of CCs, i.e. characterized by relative low entropy gas (albeit not as low as in CC systems) and a metal abundance excess.
We note that we did not use the minimum entropy \cite[e.g.][]{Cavagnolo:2009} to characterize our systems, but the ratio of entropy in the core to entropy in an outer region, this estimator is less sensitive to details of the entropy distribution and to angular resolution issues, a point later picked by other authors \citep{Hogan:2017}.
The specific value of the entropy ratio used to discriminate between CC and NCC,
was chosen after a detailed analysis of X-ray, optical and radio properties
of the objects in our sample, see \cite{Leccardi:2010} for details. It should however be pointed out that, given the lack of a clear separation between CC and NCC, there is a certain amount of arbitrariness in the way they are separated from one another (we return to this point in Sect. \ref{sec:discussion}). 
  We  dubbed  CC like structures “cool core remnants”, CCR, since we
interpreted them as the remains of a Cool Core after a heating event. 
 CCRs are a subclass of NCCs, operationally they have been  defined as those systems which, within the region where the pseudo-entropy, $s$, satisfies the condition $s/s_{OUT} < 0.8$, where $s_{OUT}$  is the pseudo entropy computed in a region with bounding radii 0.05 $R_{180}$ and $ 0.2 R_{180}$, feature a metal abundance that  exceeds, at the 2$\sigma$ confidence level, the metal abundance measured in cluster outskirts, i.e. $Z=0.23Z_\odot$. 

In the sample presented in RM10, CCRs represent slightly more than half of NCC systems, 57\% (12/21), corresponding to about 34\% (12/35) of the total cluster population , while CCs 
are 40\% (14/35) of the total cluster population. 
The fact that most CCRs are found
in dynamically active objects lends further strength to the concept that mergers can disrupt CCs.  It also 
suggests that most NCC systems have gone through a CC phase, implying that, not only can mergers disrupt CCs but also that the  fraction of systems affected by this process is a substantial one.
Furthermore, cosmological simulations suggest that merger rates are sufficiently frequent that CC systems  have a high probability of experiencing a  major merger over a timescale of a few Gyr  \citep[e.g.][]{Fakhouri:2008}.

Having found several lines of evidence pointing to a scenario where CC systems can turn into NCC we are compelled to ask if the opposite may also occur, that is if NCCs can turn into CCs. This is the question we address in the next subsection.

\subsection{Evidence for NCC reverting to CC }
\label{sec:ncc_to_cc}
Spectral analysis of central regions of NCC clusters shows the timescale for NCCs to develop a CC is comparable to or longer than the Hubble time (e.g. RM10). There are however several lines of reasoning that seem to contradict these findings, let us go through them. 

An indication that NCCs may revert to a CC state faster than expected comes 
from the analysis of SZ selected samples. SPT has allowed to construct 
representative samples of clusters out to $z\sim 1.5$. From these data we see that the  CC fraction does not change significantly with time (see \citealp{Mcdonald:2013}, \citealp{Ruppin:2021} and \citealp{Bartalucci:2019}) and yet, from RM10  we have seen that the bulk of NCC systems in the local Universe have gone through a CC phase.

The question is then: how can we have a stable CC fraction if a sizable number of CCs have been turned into NCCs and NCC cooling times are of the order of the Hubble time?  The problem can be solved if, for some reason, NCCs can revert to a CC state on timescales shorter than those estimated through X-ray spectroscopy.

Another line of reasoning leading to short NCC to CC timescales goes as follows.
If we frame the question of the relation between different cluster classes  within an evolutionary scenario and assume  the ratio between clusters in one state and another does not change much with time, as indicated by analysis of the SZ sample reported above, then the relative occupation of the different states can be used to estimate the timescales over which objects move from one another.
In \citet{Rossetti:2011} we divided a representative sample  of clusters  into ``Radio Halo" (RH) and ``Radio Quiet" (RQ) subclasses, we further divided the RQ class into  RQ,CC and RQ,NCC, i.e. RQ systems which feature CCs and that do not.
Without going into too many details, see \citet{Rossetti:2011} for a thorough discussion, we can say that, 
the ratio  N$_{RQ}$/N$_{RH}$ over the ratio N$_{RQ,CC}$/N$_{RQ,NCC}$ provides a rough estimate for the ratio of the timescale in which NCC turn into CC, t$_{NCC->CC}$, over the timescale in which RH turn into RQ, t$_{RH->RQ}$. Applying this to the GMRT sample presented in \citet{Rossetti:2011} we estimated t$_{NCC->CC}$ / t$_{RH->RQ} = 1.7$, which for a typical RH lifetime of 1 Gyr leads to t$_{NCC->CC}\sim 1.7$ Gyr.

\section{Underlying physical processes}\label{sec:phys}

In Sect.\ref{sec:ncc_to_cc} we  presented evidence suggesting that  NCCs revert to a CC
state on  timescales of a few Gyr or less, however estimate of cooling times from NCC spectra shows these should be significantly longer (e.g. RM10). How can we resolve this  contradiction? Given the  more straightforward way in which evidence in favor of the longer timescale is recovered, we might be tempted to dismiss the alternative, however deriving cooling times from X-ray spectroscopy is not as simple as might seem. The gas in the cores of NCC systems is likely characterized by a certain degree of multi-phaseness. From work presented in RM10 we know the bulk of these systems hosts CCR. If the CC disruption process is characterized by partial thermalization we could have gas with temperatures from a fraction of a keV to several keV.
The question is if conduction  can be inhibited over timescales of the Gyr. We review the evidence below.
 
A first indication that conduction could be inhibited in clusters came from Cold Fronts. Several years ago \citet{Ettori_conduc:2000} showed that the
observed sharp temperature gradient at the discontinuity surface of cold fronts requires that  conductivity 
be reduced at least by a factor of 200, with respect to Spitzer conductivity.
Other evidence comes from the Coronae identified in the cores of several NCC clusters \citep[e.g.][]{Vikhlinin3:2001,Sun:2007}. These structures, which feature temperatures of 1-2 keV, must be effectively insulated from the hot ICM surrounding them. 
Further examples are the trailing tails found in  A2142 \citep{Eckert:2014} and Hydra A \citep{Degrandi:2016}, where the suppression factor between the gas in the tail and the ambient ICM is in the order of 1000. Interestingly, in the case of A2142, from the size of the tail and the velocity of the infalling substructure, we estimate the stripped gas has been surviving in the presence of the hot ICM for at least 600 Myr.
For Cold Fronts, Coronae and trailing tails the argument could be made that these are special regions within the ICM and that the suppression is associated to magnetic draping of a weakly magnetized plasma moving through the ICM.
However, more recent work on A2142 \citep{Eckert:2017} finds evidence of inhibited conduction 
in the region of the tail downstream from where it is being disrupted. This shows that suppression
persists after linear structures are broken down and mixing, at least at the macroscopic level, is taking place. Further evidence comes from the flat density power spectrum measured in the disrupted tail, indeed, as discussed in \citet{Gaspari:2014}, Spitzer-like conduction should wash out small-scale perturbations in the ICM, leading to a steep density power spectrum. Instead, a Kolmogorov power spectrum is often found for density fluctuations in such studies, implying $\sim 10^3$ suppression factors of the plasma conductivity.

\begin{table}
	\centering
	\caption{Sample}
	\begin{tabular}{|c|c|c|}
		\hline
		\hline
		Name      & N$_{\rm H}^{\rm a}$& $z$    \\
		\hline
		          & $10^{20}$cm$^{-2}$ &     \\
		\hline
        A1644    &   4.17   & 0.0474   \\
        A1650    &   1.28   & 0.0838   \\
        A1689    &   1.75   & 0.1824   \\
        A2256    &   4.31   & 0.0579   \\
        A3558    &   3.66   & 0.0477  \\
        A3562    &   3.55   & 0.0492  \\
        A3571    &   3.88   & 0.0390  \\
        A3667    &   4.25   & 0.0557  \\
        A4038    &   1.40   & 0.0299  \\
        A576     &   5.33   & 0.0382  \\
        A754     &   4.96   & 0.0543  \\
        MKW3s    &   2.80   & 0.0447  \\
		\hline
		\hline
	\end{tabular}
	\begin{list}{}{}
		\item[Notes:]
		$\mathrm{^{(a)}}$ Weighted average N$_{\rm H}$ from \cite{HI4PI:2016}
	\end{list}
	\label{tab:sample}
\end{table}

\begin{table*}
	\centering
	\caption{Sample and spectral analysis results.}
	\begin{tabular}{|c|c c c|c c c|c c c|}
		\hline
		\hline
		Name          & \multicolumn{3}{|c|}{$N^{\rm a}$} & \multicolumn{3}{|c|}{$Z$}& \multicolumn{3}{|c|}{$T_{max}$-$\beta$}  \\
		\hline
		              &  \multicolumn{3}{|c|}{${10^{-17} \over 4\pi [D_A (1+z)]^2} \int n_e n_H dV $} & \multicolumn{3}{|c|}{Solar Units} 	 & \multicolumn{3}{|c|}{keV  $\,\,$                - }        \\
		\hline
		              &  $\alpha=0.5$ &  $\alpha=0.99$ & $\alpha=2$ &  $\alpha=0.5$ &  $\alpha=0.99$ & $\alpha=2$ & $\alpha=0.5$ & $\alpha=0.99$ & $\alpha=2$  \\
	                  &     p=-2      &    p=-100   &   p=1       &  p=-2         &  p=-100      &   p=1      &    p=-2      &  p=-100     &   p=1       \\
		\hline
        A1644        & 3.82 & 3.80 & 3.75 & 0.62 & 0.63 & 0.66 & 6.58  - 0.221 & 6.04  - 0.230 & 5.13  - 0.227 \\
        A1650        & 9.02 & 9.02 & 9.03 & 0.41 & 0.41 & 0.40 & 7.06  - 0.491 & 7.01  - 0.479 & 6.88  - 0.427 \\
        A1689        & 9.44 & 9.45 & 9.33 & 0.33 & 0.33 & 0.33 & 20.0 -  0.153 & 18.6  - 0.131 & 14.5  - 0.010 \\
        A2256        & 5.83 & 5.83 & 5.80 & 0.34 & 0.34 & 0.34 & 8.74  - 0.230 & 8.16  - 0.228 & 6.95  - 0.211 \\
        A3558        & 15.6 & 15.6 & 15.6 & 0.35 & 0.35 & 0.34 & 9.30  - 0.246 & 8.80  - 0.238 & 7.65  - 0.196 \\
        A3562        & 5.14 & 5.14 & 5.13 & 0.44 & 0.44 & 0.44 & 6.33  - 0.397 & 6.13  - 0.401 & 5.69  - 0.403 \\
        A3571        & 42.8 & 42.8 & 42.8 & 0.39 & 0.39 & 0.38 & 12.2  - 0.237 & 11.3  - 0.235 & 9.86  - 0.179 \\
        A3667        & 8.34 & 8.32 & 8.25 & 0.30 & 0.30 & 0.31 & 9.46  - 0.191 & 8.60  - 0.195 & 7.11  - 0.117 \\
        A4038        & 23.9 & 23.9 & 23.9 & 0.35 & 0.35 & 0.35 & 4.50  - 0.424 & 4.37  - 0.428 & 4.10  - 0.433 \\
        A576         & 4.68 & 4.68 & 4.67 & 0.45 & 0.45 & 0.45 & 6.20  - 0.318 & 5.94  - 0.317 & 5.36  - 0.305 \\
        A754         & 16.2 & 16.1 & 16.0 & 0.33 & 0.33 & 0.33 & 14.3 -  0.195 & 12.6  - 0.215 & 11.5  - 0.010 \\
        MKW3s        & 15.7 & 15.7 & 15.7 & 0.41 & 0.41 & 0.41 & 5.17  - 0.368 & 4.98  - 0.372 & 4.58  - 0.375 \\
		\hline
		\hline
	\end{tabular}
\begin{list}{}{}
	\item[Notes:]
	$\mathrm{^{(a)}}$ The normalization is integrated over all temperatures between $T_{min}$ and $T_{max}$.
\end{list}
	\label{tab:spec_ana}
\end{table*}

\section{Constraints from spectral analysis} \label{sec:con_spec}
Having established that conduction can be inhibited in the ICM, we investigate the limits that can be imposed on the multi temperature structure with currently available X-ray data.
We do this by performing a spectral  analysis of  CCRs, that is NCC systems featuring a central region that is likely  the remnant of a disrupted CC. We specifically target CCRs to investigate if the low entropy gas originally residing in CsC before their disruption has been at least in part retained.  
Unfortunately, NCC systems feature flat central surface brightness profiles which implies that the resolving power of XMM-Newton gratings cannot be brought to bear on these systems and that we are limited to the resolution afforded by CCD detectors. As we shall see in the next paragraphs, while less then optimal, constraints from the EPIC CCDs can be of some use.

We fit our spectral data with an emission model that assumes a powerlaw distribution of the different phases that contribute to the spectrum.  This is a somewhat rough approximation; however it does constitute a major step forward with respect to simple one temperature models that we, and others, have adopted in the past. To derive constraints on the thermodynamic quantities we shall also assume that the different gas phases in the core are in pressure equilibrium with each other.

\subsection{The emission model}
\label{sec:em_model}
We shall assume a differential emission model of the form:

\begin{equation}
	dEM = EM_* \Big( {T\over T_{max}}\Big)^{\alpha-1} {dT \over T_{max}} \, ,
\label{eq:dem}
\end{equation}

where $dEM$ is the differential emission measure associated to the plasma at temperature $T$, 
$T_{max}$ is the maximum temperature, $\alpha$ parameterizes the slope of distribution and $EM_*$ is the normalization. It is easy to show 
that 
\begin{equation}
	EM_* = \alpha EM  \, ,
	\label{eq:em_norm}
\end{equation}
where $EM$ is the integral of $dEM$ from 0 to  $T_{max}$.
The emission measure of the model we shall adopt is the integral of (1) extended from the minimum temperature, $T_{min}$, 
to the maximum temperature $T_{max}$, i.e.:

\begin{equation}
	EM(T_{min},T_{max}) = EM_* \int_{T_{min}}^{T_{max}}  \Big( {T\over T_{max}}\Big)^{\alpha-1} {dT \over T_{max}} \, .
\label{eq:em}
\end{equation}

The free parameters of the model are: $T_{min}$, $T_{max}$, the metal abundance $Z$, which we assume to be the same for all temperatures
and the normalization, $EM(T_{min},T_{max})$. 
Although $\alpha$ could be treated as a free parameter, given the strong degeneracy with the other spectral parameters, i.e. the maximum and particularly the minimum temperature, we prefer to fix it to 3 different values namely:
0.5, 0.99 and 2 and run three sets of spectral fits, one for each value of $\alpha$. The $\alpha = 0.99$ value samples the case of a flat differential emission profile, $\alpha = 0.5$ and
$\alpha = 2$ values represent the case of steep differential emission profiles, respectively increasing and decreasing with temperature (see Eq.\ref{eq:dem}). Values of $\alpha$ smaller than 0.5 or larger than 2 would be even steeper and closer to the single temperature model used in the past. 

\subsection{The isobaric model}
\label{sec:iso_model}
Our emission model only assumes the differential emission has a power-law dependency on the temperature, nothing more.
However, if we wish to derive estimates of the entropy and cooling time of the plasma responsible for the emission, we need to add further constraints. We assume that a standard equation of state applies, i.e. $p_{g}= n_{g}kT$, where $p_{g}$ is the gas pressure and $n_{g}$ the gas density, and that the different phases are in pressure equilibrium with each other, i.e. $p_{g} = $const. 

As already pointed out, NCC systems  do not feature the steep central pressure gradients found in CC systems \citep[e.g. Fig.2 of][]{Arnaud:2010}. 
Under these circumstances isobaricity, at least on the large scales we address here\footnote{On smaller scales turbulence can induce significant pressure fluctuations, see \cite{Gaspari:2014}.}, is a reasonable approximation.  

From the conditions described above, we can readily derive the thermodynamic variables describing the emitting plasma, see App.\ref{sec:app} for details. Here we recall the most important steps. By combining the best fitting emission measure, $EM(T_{min},T_{max})$ with an estimate of the volume of the emitting region, $V_{tot}$, we derive and estimate of the mean electron density, and, with some algebra, of the electron density of the hottest phase,  $n_{e,o}$.  From this, again with some algebra,  we derive an expression for the gas mass as a function of entropy\footnote{Following the standard approach in ICM studies, we define the ``entropy" $K$ as $K\equiv T/n_e^{2/3}$, where $n_e$ is the electron density.} bound between the 
minimum entropy $K_{min}$ associated to $T_{min}$ and entropy $K$, $M_{gas}(K>K_{min})$,

\begin{multline}
		M_{gas}(K>K_{min}) = \mu_e m_p n_{e,o} V_o {\alpha + 2 \over \alpha +1 } \\ 
	\Biggl[ \Biggl( {K \over K_o}\Biggr)^{{3\over 5}(\alpha +1)} - \Biggl( {K_{min} \over K_o}\Biggr)^{{3\over 5}(\alpha +1)}\Biggr]  \, ,
    \label{eq:m_vs_k}
\end{multline}

where $\mu_e=1.12 $ is the  mean molecular weight per free electron, $m_p$ is the proton mass, $V_o$ is the volume containing all phases from entropy 0 to $K_o$ and $K_o$ is the entropy of the hottest and least dense phase, i.e. $K_o = T_{max} / n_{e,o}^{2/3}$.
 
\begin{figure}
	\centerline{\includegraphics[angle=-90,width=8.8cm]{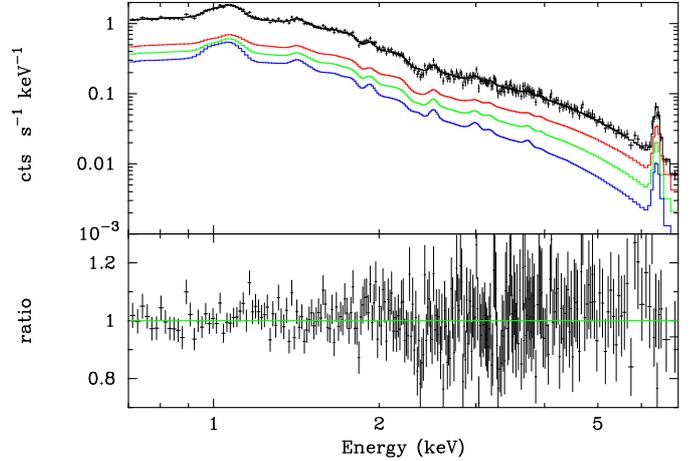}}
	\caption{Spectrum extracted from the low entropy region of MKW3s fit with a sum of 3 single temperature {\tt apec} models approximating the best-fitting {\tt wdem} model (see text for details). In the top panel we show the observed spectrum, the best fitting model for the $p = -100$ ($\alpha= 0.99$) case and its breakdown into 3 components,  indicated respectively in red (4.1 keV), green (3.1 keV)  and blue (2.4 keV). In the bottom panel  we show residuals in the form of a ratio of data over model.}
	\label{fig:mkw3s}
\end{figure}

 \subsection{Data set and results from spectral analysis}
 \label{sec:res_spec}
 We applied the emission model to a subsample (see Table \ref{tab:sample}) of the non-CC sample described in RM10. More specifically we considered all systems that were identified as CCRs. We specifically targeted CCRs  to investigate if the low entropy gas originally residing in CCs before their disruption has been at least in part retained.  
 
 We use spectra extracted from the low entropy regions found in these systems. The regions were identified through pseudo-entropy maps and feature a typical extraction radius of about 120 kpc (see RM10 for images, extraction regions and other details). Although the data reduction is the same reported in RM10, the spectra were re-extracted using a more recent version of the SAS analysis system, namely SAS16.1.
  The actual model used within  {\tt XSPEC}v12 \citep{Arnaud_XSPEC:1996} to fit the data is {\tt wdem}. While equivalent to the model described in Sect.\ref{sec:em_model} {\tt wdem} parameters are not always the same, more specifically
 the slope is defined through $p$ where $p = 1/(\alpha -1)$ and the temperature range through $\beta$, where $\beta = T_{min}/T_{max}$. Other parameters are the maximum temperature $T_{max}$, the metal abundance $Z$, which we have set to \cite{AG:1989} units, and the switch which is set to 2, requiring that the multi temperature model be constructed as the sum of {\tt apec} single temperature models.
  
 To help readers gain insight into what drives the best fit, 
 in Fig.\ref{fig:mkw3s} we show a fit to MKW3s with a sum of 3 {\tt apec} models that approximates reasonably well the best fitting {\tt wdem} model.  They are indicated respectively in red (4.1 keV), green (3.1 keV)  and blue (2.4 keV). It is important to note how, as we move from high to low temperature the contribution to the Fe L-shell bump becomes more prominent and gradually shifts to lower energies. Inclusion of cooler components would result in an even stronger and more pronounced low-energy L-shell blend, leading to increasing tension with the data. Indeed it is from the comparison of the observed emission around 1 keV with the modeled one that a minimum temperature is derived.
 
 The relative intensity between the three components is determined by the choice of $p$ ($\alpha$) which, in this example, has been set to -100 (0.99). As already mentioned, we have limited our exploration to $p=$-2,-100 and 1 ($\alpha=$0.5,0.99,2). 
 
 Since we do not deproject our spectra, we are in principle susceptible to contamination from cool gas along the line of sight. There are several reasons why we expect this contamination to be modest or even negligible.
 As shown in Table \ref{tab:spec_ana}, the ratio between the maximum and the minimum temperature is in all cases larger than 2 and in most larger than 3; if we consider typical ICM temperature profiles \citep[e.g.][]{Ghirardini:2019} we find that contributions from  2, 3 times cooler gas come from cluster outskirts, beyond R$_{500}$ or even R$_{200}$ where the emission measure is at least 2.5, 3 orders of magnitudes smaller than at the  center. Furthermore, the contribution from cooler components makes itself felt mainly through the Fe L-shell line complex, see Fig.\ref{fig:mkw3s} and related discussion, thus it will be further reduced in the metal poor gas \citep[e.g.][]{Ghizzardi:2021} found in the outskirts.
 The last and possibly strongest argument against substantial contamination from cool gas along the line of sight will be presented in Sect.\ref{sec:res_iso}.
  
Best fitting parameters from the analysis of our systems are reported in Table \ref{tab:spec_ana}. Variations in best fit parameters between the different $p$ values are of the same order if not larger than errors on the individual fits, for this reason we refrain from including errors in Table \ref{tab:spec_ana}. In systems such as MKW3s a minimum temperature can be derived rather robustly. As already discussed, this is achieved mostly  through the Fe L-shell blend. However, for others, as A754, where such a feature is not as prominent, contributions from very low temperature gas is consistent with the data and the $\beta$ parameter, see Table  \ref{tab:spec_ana}, may reach very small values.
 
\begin{table}
	\centering
	\caption{$\chi^2$ and degrees of freedom for single and multi temperature fits.}
     \resizebox{9cm}{!}{	
     	\begin{tabular}{|c|r r |r r|r r|r r|}
		\hline
        Name      & \multicolumn{2}{|c|}{1T} & \multicolumn{2}{|c|}{$\alpha=0.5$} & \multicolumn{2}{|c|}{$\alpha=0.99$}  & \multicolumn{2}{|c|}{$\alpha=2$}  \\
                  &               &          & \multicolumn{2}{|c|}{p=-2} & \multicolumn{2}{|c|}{p=-100}  & \multicolumn{2}{|c|}{p=1}   \\
		\hline
                  & $\chi^2$ & d.o.f. & $\chi^2$ & d.o.f. & $\chi^2$ & d.o.f. & $\chi^2$ & d.o.f. \\ 
		\hline
		A1644     &  434  &  409 &  405 &  408  &  405 &  408 &  405 &  408 \\ 
		A1650     & 1047  & 1017 & 1041 & 1016  & 1041 & 1016 & 1041 & 1016 \\ 
		A1689     & 1133  & 1076 & 1101 & 1075  & 1099 & 1075 & 1099 & 1075 \\ 
		A2256     &  420  &  492 &  415 &  491  &  415 &  491 &  415 &  491 \\ 
		A3558     & 1545  & 1247 & 1460 & 1246  & 1458 & 1246 & 1456 & 1246 \\ 
		A3562     &  863  &  819 &  856 &  818  &  856 &  818 &  856 &  818 \\ 
		A3571     & 1532  & 1317 & 1491 & 1316  & 1490 & 1316 & 1485 & 1316 \\ 
		A3667     & 1415  & 1138 & 1321 & 1137  & 1322 & 1137 & 1327 & 1137 \\ 
		A4038     & 1338  & 1107 & 1294 & 1106  & 1294 & 1106 & 1296 & 1106 \\ 
		A576      &  461  &  550 &  455 &  549  &  455 &  549 &  455 &  549 \\ 
		A754      &  974  & 1001 &  961 & 1000  &  960 & 1000 &  940 & 1000 \\ 
		MKW3s     & 1286  & 1061 & 1225 & 1060  & 1226 & 1060 & 1229 & 1060 \\ 
		\hline
	\end{tabular}}
	\label{tab:chi2}
\end{table}

 It is worth pointing out that in the vast majority of cases our multi temperature model provides a better fit than a single temperature model. As shown in Table \ref{tab:chi2}, in all instances there is a  reduction in $\chi^2$, when going from the one-temperature to the multi-temperature fits. By applying an F-test we find the improvement to be significant at more than the 99\% level in all but two cases, namely A1650 and A2256 where significance is at the 98\% level, see Table \ref{tab:ftest}. 
 \begin{table}
 	\centering
 	\caption{F statistic value and associated probability of chance improvement obtained by comparing multi temperature with single temperature $\chi^2$ and d.o.f.}
 	\resizebox{9cm}{!}{	
 		\begin{tabular}{|c|r l|r l|r l|}
 			\hline
 			Name    & \multicolumn{2}{|c|}{$\alpha=0.5$} & \multicolumn{2}{|c|}{$\alpha=0.99$}  & \multicolumn{2}{|c|}{$\alpha=2$}  \\
 			        & \multicolumn{2}{|c|}{p=-2} & \multicolumn{2}{|c|}{p=-100}  & \multicolumn{2}{|c|}{p=1}   \\
 			\hline
 			          & F-stat & Prob. & F-stat & Prob. & F-stat & Prob. \\ 
 			\hline
 			A1644     & 29.2 & $1.1 \, 10^{-7}$  & 29.2 & $1.1 \, 10^{-7}$  & 29.2 & $1.1 \, 10^{-7}$  \\ 
 			A1650     & 5.86 & $1.6 \, 10^{-2}$  & 5.86 & $1.6 \, 10^{-2}$  & 5.86 & $1.6 \, 10^{-2}$  \\ 
 			A1689     & 31.2 & $2.9 \, 10^{-8}$  & 33.3 & $1.1 \, 10^{-8}$  & 33.3 & $1.1 \, 10^{-8}$  \\ 
 			A2256     & 5.92 & $1.5 \, 10^{-2}$  & 5.92 & $1.5 \, 10^{-2}$  & 5.92 & $1.5 \, 10^{-2}$  \\ 
 			A3558     & 72.5 & $4.7 \, 10^{-17}$ & 74.3 & $2.0 \, 10^{-17}$ & 76.2 & $8.2 \, 10^{-18}$ \\ 
 			A3562     & 6.69 & $9.8 \, 10^{-3}$  & 6.69 & $9.8 \, 10^{-3}$  & 6.69 & $9.8 \, 10^{-3}$  \\ 
 			A3571     & 36.2 & $2.3 \, 10^{-9}$  & 37.1 & $1.5 \, 10^{-9}$  & 41.7 & $1.5 \, 10^{-10}$ \\ 
 			A3667     & 80.9 & $9.7 \, 10^{-19}$ & 80.0 & $1.5 \, 10^{-18}$ & 75.4 & $1.3 \, 10^{-17}$ \\ 
 			A4038     & 37.6 & $1.2 \, 10^{-9}$  & 37.6 & $1.2 \, 10^{-9}$  & 35.8 & $2.9 \, 10^{-9}$  \\ 
 			A576      & 7.24 & $7.4 \, 10^{-3}$  & 7.24 & $7.4 \, 10^{-3}$  & 7.24 & $7.4 \, 10^{-3}$  \\ 
 			A754      & 13.5 & $2.5 \, 10^{-4}$  & 14.6 & $1.4 \, 10^{-4}$  & 36.2 & $2.5 \, 10^{-9}$  \\ 
 			MKW3s     & 52.8 & $7.2 \, 10^{-13}$ & 51.9 & $1.1 \, 10^{-12}$ & 49.2 & $4.2 \, 10^{-12}$ \\ 
 			\hline
 	\end{tabular}}
 	\label{tab:ftest}
 \end{table}

  \begin{figure}
 	\centerline{\includegraphics[angle=-90,width=8.8cm]{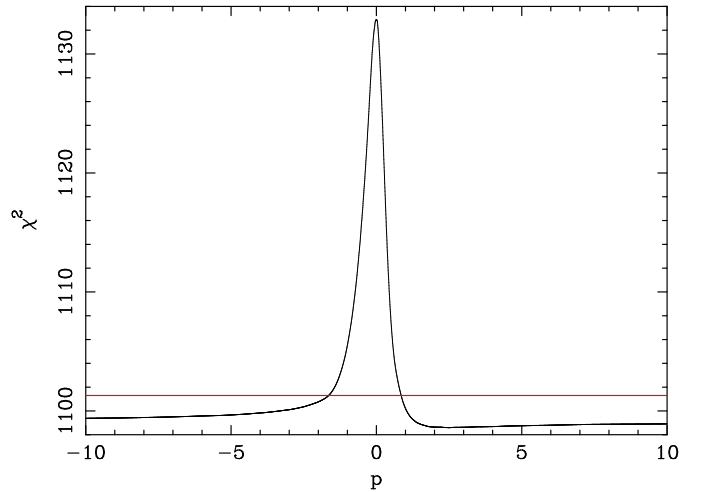}}
 	\caption{Confidence region for the slope parameter p for the case of A1689, the red line correspond to 90\% confidence.
 	Note how only $\rm p$ values close to 0 (single temperature model) can be rejected at high significance.}
 	\label{fig:a1689cont}
 \end{figure}

 Having presented our spectral results, we review ex post facto our choice of working with fixed values of the slope $\rm $p ($\alpha$) rather than allowing it vary freely. In most cases exploratory fits showed a strong degeneracy between the slope and other parameters, mainly $\beta$ (the ratio of minimum to maximum temperature).  What we discovered is that, while 
 EPIC spectra have sufficient resolution to discriminate between single and multi-temperature models (see Tables \ref{tab:chi2}, \ref{tab:ftest} and associated discussion) they are relatively insensitive to the details of the multi-temperature model. As an example we consider the case of A1689; in Fig.\ref{fig:a1689cont} we show confidence regions for parameter p. As we can see,  the $\chi^2$ changes only moderately as we go from $\rm p =-10$ to $\rm p \sim -2$, where it starts rising reaching a maximum at  $\rm p = 0$, it then falls rapidly until  $\rm p \sim 1$, and from here stays relatively flat all the way up to $\rm p = 10$. It is also worth mentioning that, although not shown in Fig.\ref{fig:a1689cont},   $\chi^2$ values for  $\rm p < -10$ and  $\rm p > 10$ are bracketed between 1099 and 1100. Interestingly,  $\chi^2$ values for 
 $\rm p =-2,-100,1$ ($\alpha=0.5,0.99,2$) are all within the 90\% confidence region. Furthermore, only fits in the range $-0.78< \rm p <0.47$ ($\alpha < -0.28$ and $\alpha > 3.1$), which approximate the single temperature case, can be rejected with a high statistical significance ($>99.9$\%). 
 By inspecting  Table \ref{tab:chi2} we find that in most instances $\chi^2$ values for fits with $\rm p = -2,-100,1$ are very similar, as found for A1689. Moreover we see that the minimum value is not always associated
 to a specific value of $\rm p$, but changes from cluster to cluster. This tells us that, for most of our systems, spectral fits do not show a preference for a specific value of $\rm p$. Our choice of the 3 values $\rm p =-2,-100,1$, corresponding to $\alpha=0.5,0.99,2$ can be justified in retrospect by noting that $\rm p=-2$ $(\alpha=0.5)$ and $\rm p = 1$ $(\alpha=2)$
 sample  2 cases, the first with differential emission measure decreasing with increasing temperature and the second with differential emission measure increasing with increasing temperature (see Eq.\ref{eq:dem}), that while relatively near to the single temperature case ($\rm p = 0$, $\alpha=\infty$),  are sufficiently distant to afford  an acceptable fit. Conversely,  $\rm p = -100$ $(\alpha=0.99)$, samples the case of a flat differential emission measure.

 Our multi-temperature best fits allow for a significant amount of low temperature gas in most of the systems.  However we must be cautious not to over interpret results, indeed the contribution of the low temperature gas to the spectra is modest and alternative fits of similar quality can be obtained with 2 temperature models that do not include as much cool gas. In simpler words we can say that our data is consistent with the presence of low temperature gas but does not require it. As we shall see in Sect.\ref{sec:future}, direct evidence of low temperature gas can only be achieved with high spectral resolution instruments capable of resolving out emission lines from the low temperature gas.

\subsection{ Results from application of the isobaric model}
\label{sec:res_iso}

\begin{figure}
	\centerline{\includegraphics[angle=0,width=8.8cm]{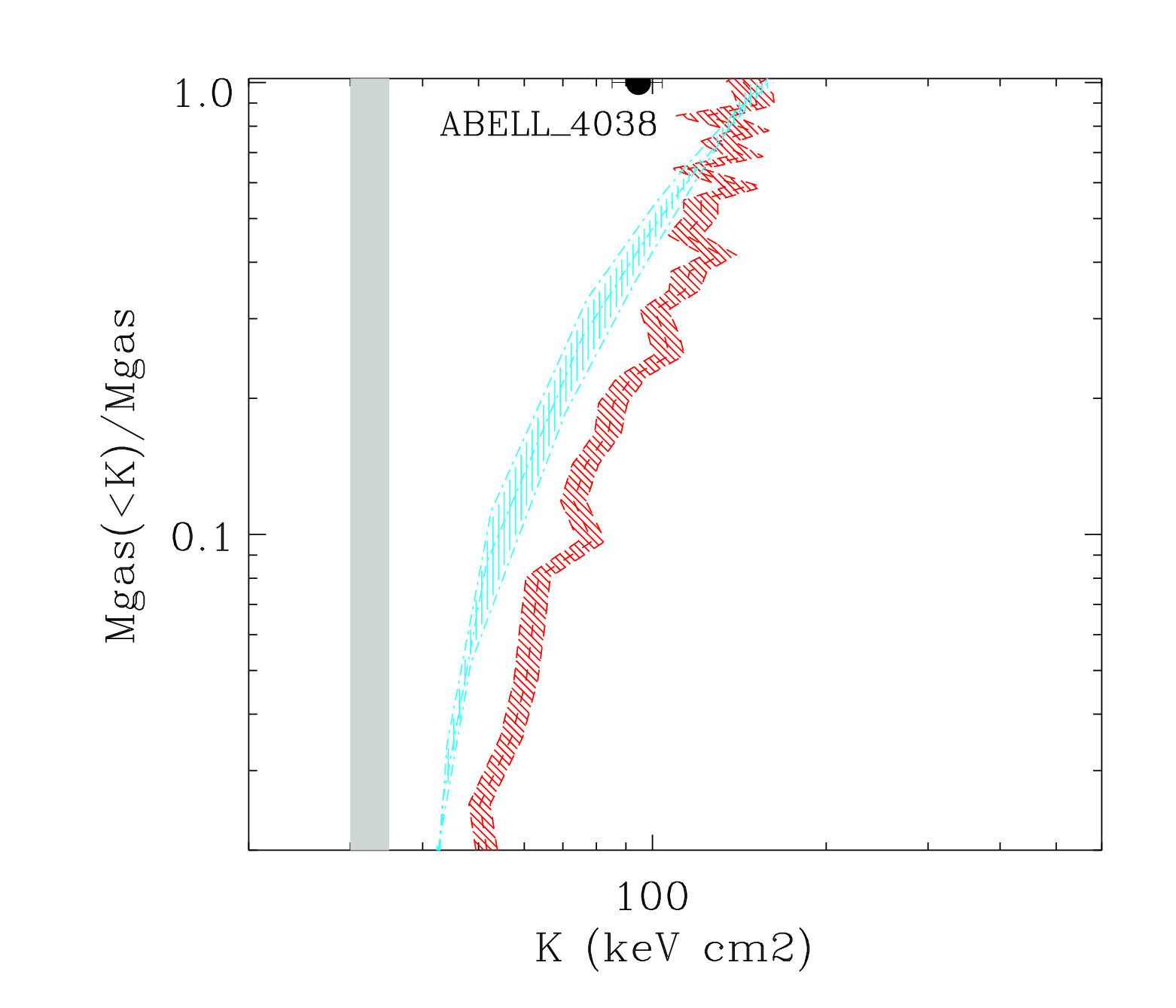}}
	\centerline{\includegraphics[angle=0,width=8.8cm]{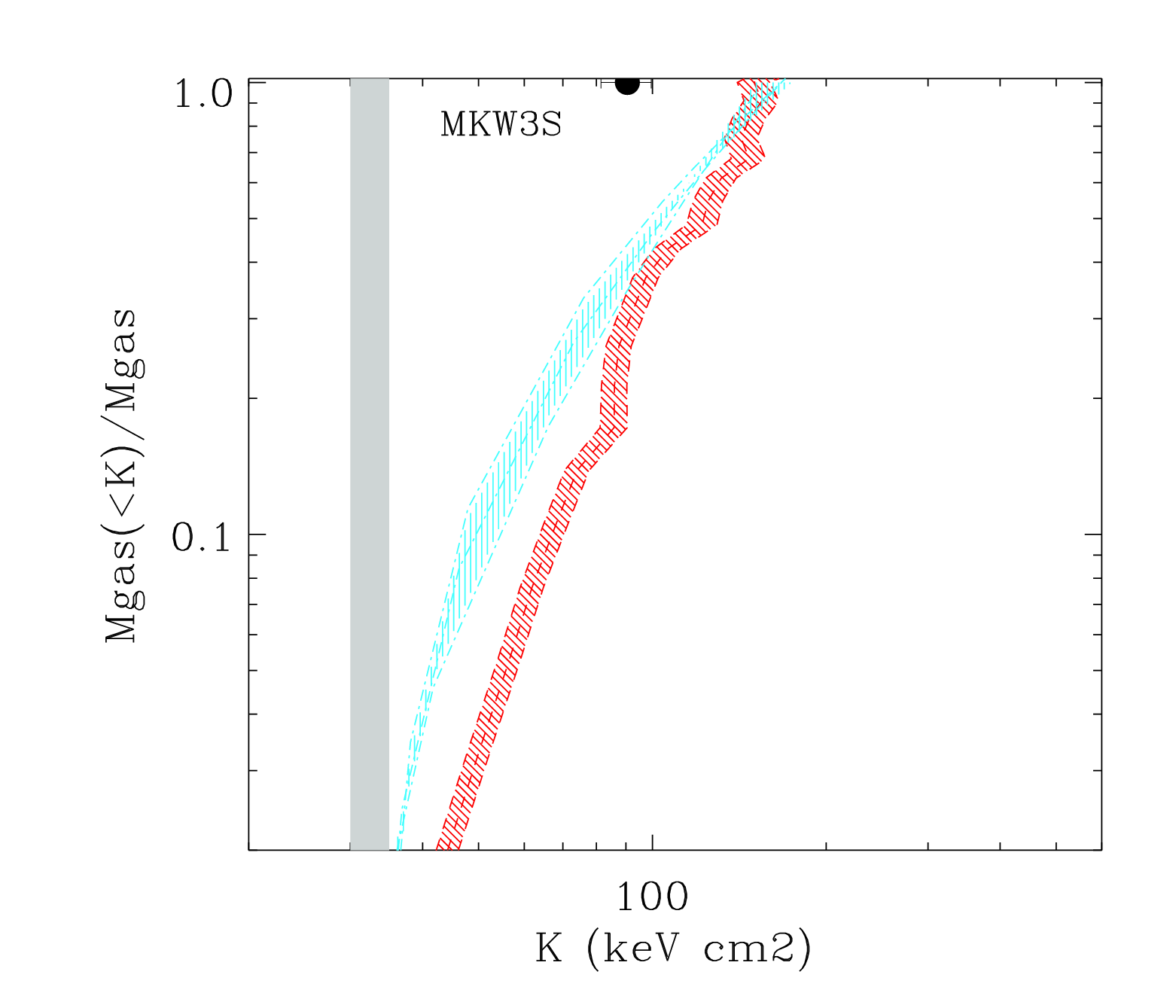}}
	\caption{Normalized cumulative gas mass versus entropy as derived with the isobaric model (light blue shaded region) and single temperature modeling of spatially resolved spectra from the ACCEPT archive of Chandra data (red shaded region). The distributions are normalized by dividing by the total gas mass. We show, as a filled circle  the value of the entropy obtained from a single temperature analysis of our spectra. The gray vertical box corresponds to $K=30-33$ keV cm$^2$. Note that only CC objects are expected to have mass distributions extending below this entropy. Data for A4038 and MKW3s are shown respectively in the top and bottom panels.}
	\label{fig:with_accept_m_int_vs_k}
\end{figure}
Before proceeding with the description of our results we present a sanity check we performed on 2 of our systems, namely A4038 and  MKW3s. For these objects we have used high spatial resolution spectral results obtained with Chandra and made  available through the ACCEPT database \citep{Cavagnolo:2009}. From the ACCEPT radial profiles of gas density and entropy we have reconstructed the normalized cumulative gas mass versus entropy profiles. The gas mass is normalized to the mass enclosed within $\sim$ 120 kpc, which is comparable to the size of the region from which our CCR spectra have been extracted. Among our systems, A4038 and  MKW3s are those for which the ACCEPT archive finds the strongest entropy gradient and therefore the ones that can be better characterized through high angular resolution spectral analysis. We have compared gas mass versus entropy distributions obtained with our isobaric model with those obtained from ACCEPT through spatially resolved spectroscopy.  Comparisons are reported in Fig.\ref{fig:with_accept_m_int_vs_k}. As can be seen, the mass vs. entropy distributions, while not identical, are similar.
Both feature a sizable fraction of gas mass for which the entropy is substantially smaller than the one derived from a single  temperature fit, shown as a black dot in the figure. Moreover, since A4038 and  MKW3s contain the strongest pressure gradients among our systems, they should also be those for which our model, which is isobaric, is less precise. Note also that, since data from \cite{Cavagnolo:2009} has been deprojected, the comparison shown in Fig.\ref{fig:with_accept_m_int_vs_k} can be used to argue that the contamination from gas along the line of sight discussed in Sect.\ref{sec:res_spec}, must indeed be modest. 

\begin{figure}
	\centerline{\includegraphics[angle=0,width=8.8cm]{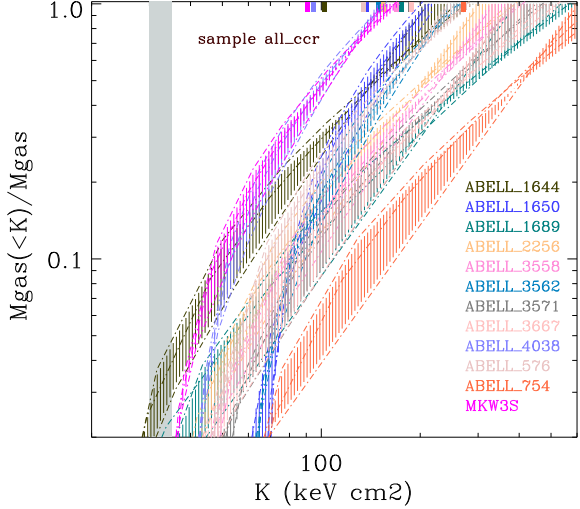}}
	\caption{Normalized cumulative gas mass versus entropy for the objects in our sample. For each object we plot the distributions for $p=-2, -100, 1$ ($\alpha=0.5,0.99,2$) and shade the region between them. The gas mass is normalized to the total gas mass.
		For comparison we show, as small rectangles following the same color coding of the gas mass distributions, the values of the entropy obtained from a single temperature analysis of our spectra. The gray vertical box corresponds to $K=30-33$ keV cm$^2$. Note that only CC objects are expected to have mass distributions extending below this entropy.}
	\label{fig:all_m_int_vs_k}
\end{figure}

\begin{table}
	\centering
	\caption{Comparison of entropy and cooling times derived from the single temperature model with those constraining 10\% of the gas mass in the isobaric model and after reconstruction of the pressure gradient.}
		\resizebox{9cm}{!}{
			\begin{tabular}{|r|r|r|r|r|r|}
		\hline
		\hline
		Name      & $K_{1T}^\mathrm{^{(a)}}$   &  $K_{0.1}^\mathrm{^{(b)}}$   & $t_{cool,1T}^\mathrm{^{(c)}}$ &  $t_{cool,0.1}^\mathrm{^{(d)}}$ &  $t_{cool,0.1,r}^\mathrm{^{(e)}}$ \\
		& keV cm$^{2}$ &   keV cm$^{2}$  & Gyr     & Gyr  & Gyr  \\
		\hline
		A1644      &   103         &  53      &   10.0    &     4.1  &   3.3      \\
		A1650      &   138         &  83      &   10.5    &     6.1  &   4.2  \\           
		A1689      &   176         &  90      &   9.3     &     4.2  &   3.7     \\
		A2256      &   155         &  79      &   13.0    &     6.1  &   4.6            \\
        A3558      &   169         &  83      &   12.7    &     6.5  &  4.9         \\
        A3562      &   149         &  81      &   13.1    &     7.2  &  5.1           \\
        A3571      &   186         &  92      &   12.4    &     6.2  &  4.7     \\
        A3667      &   188         &  91      &   17.1    &     8.0  &  6.1             \\
        A4038      &   95          &  51      &    9.5    &     5.0  &  3.3        \\
        A576       &   134         &  70      &   13.0    &     6.1  &  4.5            \\
        A754       &   271         & 131      &   20.8    &     10.1 &  8.3            \\
        MKW3s      &   91          & 48       &   8.1     &     4.1  &  2.9            \\
		\hline
		\hline
	\end{tabular}}
	\begin{list}{}{}
		\item[Notes:]
		$\mathrm{^{(a)}}$ entropy derived from the single temperature model; 
		$\mathrm{^{(b)}}$ entropy constraining 10\% of the gas mass in the isobaric model; 
		$\mathrm{^{(c)}}$ cooling time derived from the single temperature model;
		$\mathrm{^{(d)}}$ cooling time  constraining 10\% of the gas mass in the isobaric model; 
	    $\mathrm{^{(e)}}$ cooling time  constraining 10\% of the gas mass after reconstitution of the pressure gradient.
	\end{list}
	\label{tab:frac}
\end{table}

\begin{figure}
	\centerline{\includegraphics[angle=0,width=8.8cm]{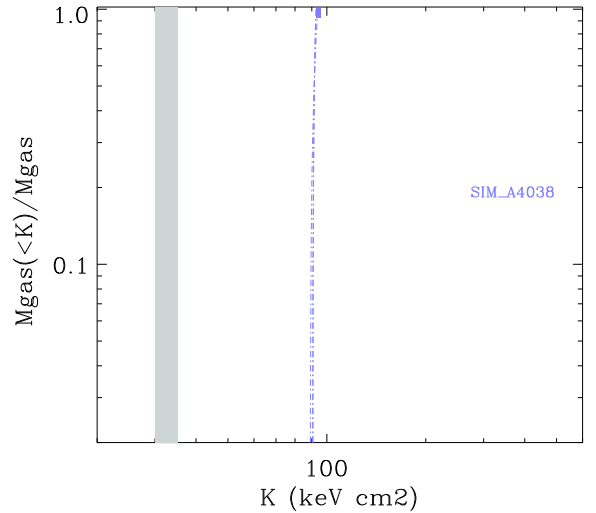}}
	\caption{Normalized cumulative gas mass versus entropy  for a spectrum simulated from the one temperature best-fit to A4038. 
		The gray vertical box corresponds to $K=30-33$ keV cm$^2$. Note that the gas mass versus entropy distribution is essentially vertical and the value of the entropy it identifies is in good agreement with the one determined from a single temperature analysis of the data.}
	\label{fig:m_int_norm_vs_k_sim}
\end{figure}

In Fig.\ref{fig:all_m_int_vs_k} we show cumulative gas mass fractions as a function of entropy for our systems. 
In all cases the assumption of a distribution rather than a single temperature leads to significant amounts of gas with entropies well below the entropy derived from the single temperature model. For example, in the case of MKW3s about 10\% of the gas has an entropy that is less than half of the single temperature model. 
For the vast majority of our objects we find that roughly 10\% of the gas mass features values of the entropy smaller than half of what is measured with the single temperature model, see Table \ref{tab:frac}.

To convince ourselves, as well as our readers, that the isobaric model works well in the limiting case of an iso-thermal gas, we have simulated a one temperature spectrum with statistics comparable to those of objects in our sample, analyzed it with the {\tt wdem}  model and computed normalized cumulative gas mass versus entropy profiles as we did for our real data. The resulting  distributions are reported in Fig.\ref{fig:m_int_norm_vs_k_sim}, as we can see the gas mass versus entropy lines are essentially vertical and the value of the entropy they identify is in good agreement with the one determined from a single temperature analysis of the data.

To better understand how the gas mass versus entropy distributions in our systems differ from those found in CC clusters we have used Chandra data made available through the ACCEPT Archive \citep{Cavagnolo:2009}. More specifically, we have taken
 12 CC systems from ACCEPT that are representative of the parent population in the sense that their entropy profiles span the range covered by CC systems. 

\begin{figure}
	\centerline{\includegraphics[angle=0,width=8.8cm]{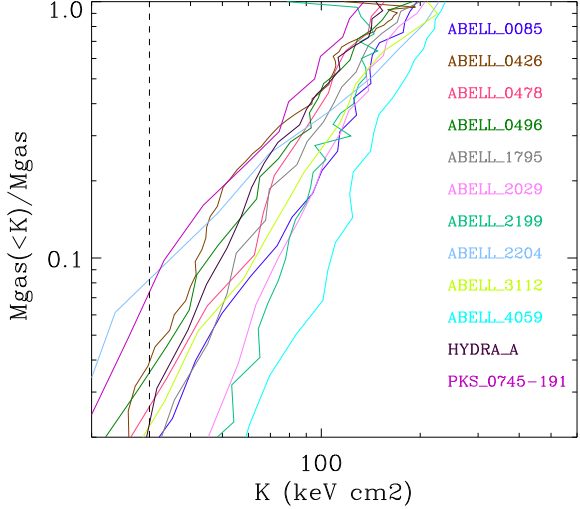}}
	\caption{Normalized cumulative gas mass versus entropy  for 12 CC systems  extracted from the ACCEPT sample. For all these systems the minimum entropy is smaller than 30 keV cm$^2$. The gas mass is normalized to the total gas mass.
		The dashed vertical line corresponds to $K=30$ keV cm$^2$. Note that only CC objects are expected to have mass distributions extending below this entropy. There are a few systems, i.e. A4059,A2199 and A2029, that have less than 1\% of their gas mass with  $K < 30$ keV cm$^2$ and which do not cross the threshold within the boundaries of this plot. }
	\label{fig:m_int_norm_vs_k_cc}
\end{figure}

As it turns out the $M_{gas}$ versus $K$ distributions of CC systems do not differ much from those of CCR systems,
for example only 4 out of 12 show an $M_{gas}(K<$30 keV cm$^2$)/$M_{gas}$ larger than 3\%.  Other CC systems show 
$M_{gas}(K<$30 keV cm$^2$)/$M_{gas}$ values comparable to those derived for CCR.
In summary, for the majority of our systems, the gas residing within $\sim$ 120 kpc 
features an entropy distribution similar to the one found in CC systems.   

\begin{figure}
	\centerline{\includegraphics[angle=0,width=8.8cm]{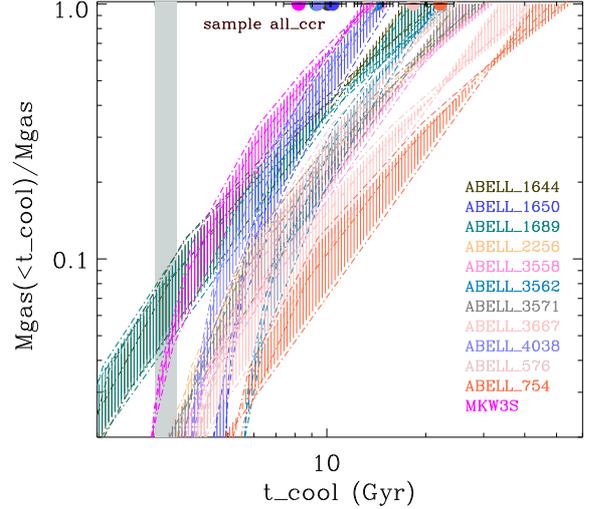}}
	\caption{Normalized cumulative gas mass versus cooling time for the objects in our sample. For each object we plot the distributions for $p=-2, -100, 1$ ($\alpha=0.5,0.99,2$) and shade the region between them. The gas mass is normalized to the total gas mass.
		For comparison we show, as small filled circles following the same color coding of the gas mass distributions, the values of the cooling time obtained from a single temperature analysis of our spectra. The gray vertical box corresponds to $t_{cool}= (3-3.3)$ Gyr. Note that only CC objects have mass distributions extending below this value.}
	\label{fig:all_m_int_vs_tcool}
\end{figure}

\begin{figure}
	\centerline{\includegraphics[angle=0,width=8.8cm]{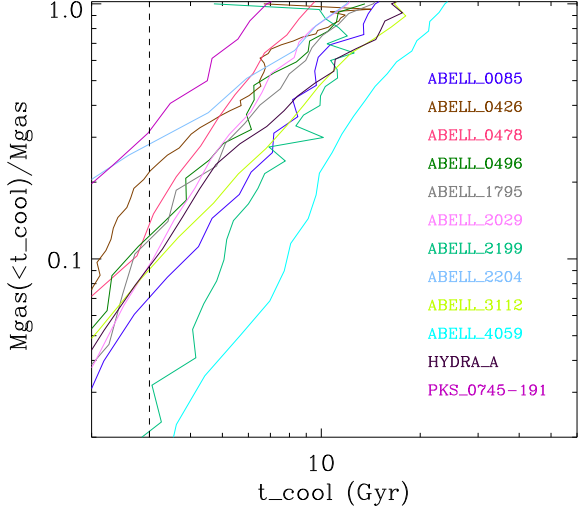}}
	\caption{Normalized cumulative gas mass versus cooling time for 12 CC systems  extracted from the ACCEPT sample. The gas mass is normalized to the total gas mass.
		The dashed vertical line corresponds to $t_{cool}= 3$ Gyr. Note that only CC objects have mass distributions extending below this time.}
	\label{fig:m_int_norm_vs_tcool_cc}
\end{figure}

Proceeding in a fashion similar to the one adopted in Sect.\ref{sec:iso_model}, we derived a relation between the gas mass and the cooling time, unfortunately since the cooling time is determined numerically we cannot provide an explicit equation, details can be found in App.\ref{sec:app}.
In Fig.\ref{fig:all_m_int_vs_tcool} we show the normalized gas mass as a function of the cooling time, for objects in our sample. 
As in the case of mass vs. entropy, the mass vs. cooling time distributions show that a large fraction of the gas features cooling times shorter than those recovered with one temperature models. 
For the vast majority of our objects we find that roughly 10\% of the gas mass features cooling times smaller than half the cooling time inferred from the single temperature model, see Table \ref{tab:frac}.
A comparison with the sample of 12 CC systems from the ACCEPT database, previously discussed, is enlightening.
As can be seen from Fig.\ref{fig:m_int_norm_vs_tcool_cc}, 4 CC system out of 12 have $M_{gas}(t_{cool}<$3 Gyr)/$M_{gas}> $ 5\%. Other CC systems feature $M_{gas}(t_{cool}<$3 Gyr)/$M_{gas}$  values comparable to those derived for most CCRs. 
In summary, under the assumptions presented in this section, a sizable fraction of the gas in the cores of our systems  has cooling times that are substantially shorter than those estimated from a single temperature analysis and not too different from those found in CC systems.

\subsection{Core reconstitution phase} \label{sec:rec}

All systems in our sample show evidence of being dynamically active (see RM10),
and they will likely evolve significantly over relatively short timescales. Thus, the question of if and how the cooling time of the gas in their cores will change over a short timescale is a meaningful one.
An important point is that, although not relaxed, 
a sizable fraction of our systems are not in the most violent merger phase, as the dearth of Radio Halos and other evidence suggests. More specifically, in RM10, we noticed that 7/12 CCR were not in a major merger phase, which was indicated by the absence of cluster-wide diffuse radio emission at 1.4 GHz, by single-peaked distributions of the galaxies' velocities, and by the lack of significant irregularities in the X-ray image and/or temperature map.

Under such conditions we expect the distribution of gravitating mass, dominated by dark matter, to tend back to equilibrium. The timescale over which this should occur is of a few crossing timescale, which for the central regions under consideration:  $R\sim 120$ kpc, and velocities typical for massive systems, $v \sim 1000$ km/s, should be smaller than $\sim$ 1 Gyr \citep[see Sect. 2.9 of][]{Sarazin:1988}. The ICM will respond to these changes by reestablishing a radial pressure gradient, note that any nonthermal pressure support should provide a modest contribution since, as already pointed out, the systems are not in the most violent merger phase. As the pressure gradient forms, convective motions in the ICM will lead to entropy stratification of the plasma. 
The velocity at which convective motions operate, a fraction of the sound speed, should be sufficient to keep up with the changes induced by the reforming potential well.
Approximating the cooling time with the expression $t_{cool} \propto T^{1/2}/n$ it is easy to recast it in terms of entropy and pressure: $t_{cool} \propto K^{9/10} p^{-2/5}$. Assuming the reconstitution process to be mostly adiabatic and considering that the lower entropy gas will experience an increase in pressure as it sinks towards the bottom of the reforming potential well, we see that the
shortest cooling times will suffer a further decrement that correlates with the increase in pressure.  
To make a more quantitative estimate of the change in cooling times in our systems, on top of  adiabatic evolution, we assume the radial profiles for the  gas pressure and entropy can be approximated by power-laws of the form: 

\begin{equation}
	p = p_o \Bigl({r \over r_{max}}\Bigr)^\epsilon \,\,\, {\rm and} \,\,\, K = K_o \Bigl({r \over r_{max}}\Bigr)^\gamma \, ,  
	\label{eq:p_k}
\end{equation}

where $p_o = n_o T_{max}$, $r_{max}$ is estimated from the size of the region from which spectra are extracted and $\epsilon$ and $\gamma$ take on values that are typical of CC systems namely:
$ \epsilon = 1$ and $\gamma = 0.7 $, see \cite{Arnaud:2010} for pressure and \cite{Panagoulia:2014} and \cite{Babyk:2018} for entropy.  With some algebra, see App.\ref{sec:appb} for details, we worked out an expression for the mass versus cooling time  that is consistent with the pressure and entropy profiles reported in Eq.\ref{eq:p_k} and with the mass vs. entropy distribution derived from the isobaric model (see Eq.\ref{eq:m_vs_k} and Fig.\ref{fig:all_m_int_vs_k}). 

\begin{figure}
	\centerline{\includegraphics[angle=0,width=8.8cm]{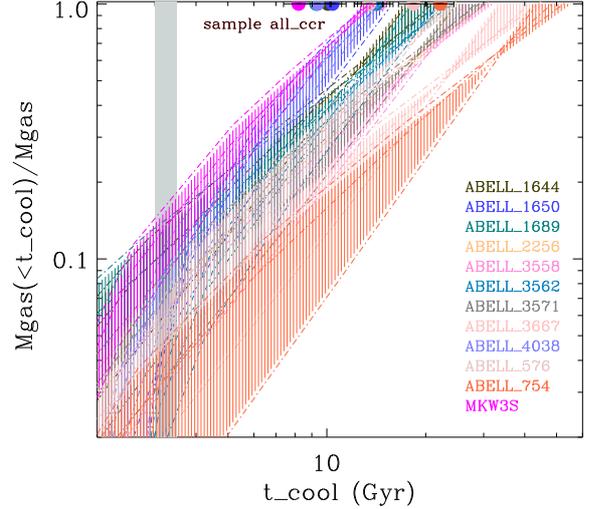}}
	\caption{Cumulative gas mass versus cooling time for the objects in our sample for the mass vs. cooling time  that is consistent with the pressure and entropy profiles reported in Eq.\ref{eq:p_k} and with the mass vs. entropy distribution for the isobaric model (Eq.\ref{eq:m_vs_k}). For each object we plot the distributions  for $p=-2, -100, 1$ ($\alpha=0.5,0.99,2$) and shade the region between them. The gas mass is normalized to the total gas mass.
		For comparison we show, as small filled circle following the same color coding of the gas mass distributions, the values of the cooling time obtained from a single temperature analysis of our spectra. The gray vertical box corresponds to $t_{cool}= (3-3.3)\,10^{9}$ Gyr. Note that only CC objects have mass distributions extending below this value.}
	\label{fig:all_m_int_vs_tcool_ref}
\end{figure}

The distribution of mass with cooling time, after the central pressure gradient is reinstated (see Fig.\ref{fig:all_m_int_vs_tcool_ref}), shows that the fraction of gas mass at the low cooling time end is significantly larger than the one derived from the isobaric model (see Fig.\ref{fig:all_m_int_vs_tcool} and Table \ref{tab:frac}). Moreover, several systems have mass versus cooling time distributions consistent with those found in CC systems (see Fig.\ref{fig:m_int_norm_vs_tcool_cc}). 
The implication is that if, after a merging event, the distribution of the gravitating mass tends back to the shape typically found in relaxed systems, the cooling time of the lower entropy gas, which will sink to the bottom and experience adiabatic compression, will be reduced to the point of being comparable to  that found in  CC systems and likely kick-start a new self-regulated AGN feedback cycle via an inner multiphase condensation and Chaotic Cold Accretion rain \citep{Gaspari:2020}.

It is worth noting that our estimate of cooling times is a conservative one. Indeed we are considering evolution of our systems over a timescale that is a non-negligible fraction of their cooling times, this implies that the process is not adiabatic and that our estimates of cooling times are actually upper limits.

\section{Discussion}\label{sec:discussion}

In this paper we have addressed the relationship between CC and NCC systems. More specifically, we have asked ourselves if objects belonging to one class can transit into the other. In Sect.\ref{sec:cc_ncc} we have reviewed 
the substantial  evidence that mergers can turn CC systems into NCCs. In Sect.\ref{sec:ncc_to_cc} we have investigated the opposite process: we have asked ourselves if central regions of NCC systems can evolve towards CC on timescales shorter than the Hubble time. There are several arguments suggesting this may be the case, the most compelling is that the number of CC systems is stable across cosmic time \citep[e.g.][]{Ruppin:2021} despite the high rate of mergers disrupting CCs on Gyr timescales.
In Sect.\ref{sec:con_spec} we have outlined a physical model of the ICM which would allow for short transition times from NCC to CC systems. The gas in the core of some NCC systems may be characterized by a distribution of phases. The lower temperature components are also the higher density ones and feature entropies and cooling times that are much smaller than those of the hotter and less dense phases. In such a system the central regions could rapidly evolve from NCC to CC. The model relies on conduction being heavily suppressed in the ICM, which is something for which we do have evidence, see Sect.\ref{sec:phys} for a thorough discussion.

The next step has been to use XMM-Newton EPIC spectra of CCRs, a subclass of NCCs, to test if  
they do allow for a small fraction of  gas with 
small entropy and cooling times possibly comparable to those found in CC systems, see Sect.\ref{sec:con_spec}.
We specifically targeted CCRs as these systems have recently been transformed from CC to NCC (see RM10) and 
might retain part of the low entropy gas originally residing in CCs.   
We have fitted our data with an emission model that assumes a powerlaw distribution of the phases contributing to the spectrum.  This is a somewhat rough approximation, however it does constitute a major step forward with respect to simple one temperature models that we, and others, have adopted in the past.
To derive constraints on the thermodynamic quantities we have also 
assumed that the different gas phases are in pressure equilibrium with each other.
In all our systems we have found that a sizable fraction of the gas can have entropy and cooling times that 
are substantially smaller than those found under the assumption of a single temperature plasma (see Figs.\ref{fig:all_m_int_vs_k} and \ref{fig:all_m_int_vs_tcool}).
 A comparison  with entropy  distributions in cores of CC systems (see Fig. \ref{fig:m_int_norm_vs_k_cc})
 shows that most CCRs host a small amount of cool gas with an entropy distribution similar to the 
 one found in  CC systems.
 For cooling times distributions the overlap between CCRs and CCs is somewhat smaller  
 (see Fig. \ref{fig:m_int_norm_vs_tcool_cc}). However under the reasonable assumption that the potential well at the center  of our systems will reform on a short timescale ($\sim$ 1 Gyr) the fall of the lower entropy gas towards the center will lead to cooling time distributions very similar to those found in CC systems (see Fig.  \ref{fig:all_m_int_vs_tcool_ref}).

An important point is that, on the basis of our estimates, the timescale over which a CC is reformed is comparable to the timescale over which mergers are active (e.g. duration of Radio Halos), this suggests that, since we observe a substantial number of systems in the merging phase we should also observe systems in the CC reconstitution phase. The question is then: where are these systems? Possibly under our very eyes. In RM10 we assumed that those CCR systems showing the smaller deviation in terms of entropy with respect to CC systems might be CCs that suffered a heating event associated to AGN feedback rather than to a merger.  An alternative explanation is that these systems, along with some of the less extreme merger CCRs, could actually be objects going through the CC reconstitution phase.  It has been pointed out to us that, if this is indeed the case, the acronym CCR should be dropped as it no longer reflects our understanding of these systems. We respond by noting that the R of CCR can stand either for "Remnant" or for "Resurgent", with the ambiguity in this term fittingly reflecting our difficulty in separating one from the other.

We now briefly discuss the implications of having low entropy gas in CCR systems.
A first consequence is that, in at least some instances, the  merging process that disrupts CCs mostly mixes the gas on the larger scales modifying only moderately or not at all the entropy of the cooler gas. Note that this does not imply the disruption process will be on the whole an adiabatic one but merely that the lower entropy gas will be, either in part or completely, preserved. 
If NCCs can be transformed into CCs on the timescale of a few Gyr, then the constant fraction of CCs over cosmic time \citep{Ruppin:2021} emerges from a dynamic equilibrium where CCs transformed into NCCs through mergers or AGN feedback are balanced by NCCs that revert to CCs. In this framework CCs and NCCs should no longer be viewed as subclasses but as ``states" between which clusters can move.  This rather general statement has some tangible implications, we now briefly discuss two of them. 

Within the framework of the evolutionary model we propose there is no net separation between CCs and NCCs as objects move from one class to another.
This naturally explains the presence of transition objects and is consistent with the fact that observational classifications based on different indicators sometimes lead to conflicting results.

Several authors \citep[e.g.][]{Fabian:2012,HL:2015} have pointed out that in CCs the energy loss from cooling is  comparable to the mechanical energy required to inflate the cavities often found in these systems. The implication being that the energy associated to the cavities can offset the cooling. If we consider CC clusters as a distinct subclass of objects, the absence of systems where the mechanical power exceeds the cooling luminosity by a large factor suggests that somehow the AGN injects just the right amount of energy
to offset the cooling. Conversely, if we allow for CCs to be transformed into NCCs such a constrain might not be required, indeed systems with high mechanical power to cooling luminosity ratio may well suffer substantial heating and transform from CC into NCC. 
Interestingly, in a recent paper \citep{Ruppin:2022}, the authors have plotted the ratio of mechanical power to cooling luminosity for CC and NCC systems finding that several NCCs show very high ratios, which is what we might expect in the scenario outlined above. 

\section{Implications for future missions}\label{sec:future}
In the previous sections we have shown that, in several CCR systems, we can have significant amounts of low entropy/cooling time gas. If this is so, it should be possible to detect it 
 through high resolution spectroscopy.
In the future we will have microcalorimeters on board XRISM \citep{Tashiro_XRISM:2018} and ATHENA \citep{Nandra_Athena:2013}.
We have carried out simulations based on spectral fits presented in the previous sections to gauge how well these instruments will characterize low temperature components. 

We have made use of XRISM 
spectral response and background files, as made available at the HEASARC website\footnote{ {\tt \small https://heasarc.gsfc.nasa.gov/docs/xrism/proposals/}}. 
We started  by performing a 500 ks simulation of A576 based on the best fitting {\tt wdem} model  for the case p=-100 ($\alpha=0.99$), see Table \ref{tab:spec_ana}. For this and all other simulations, if the region from which the spectral model  was derived exceeded the XRISM FoV we rescaled appropriately the normalization.  
\begin{figure}
	\centerline{\includegraphics[angle=-90,width=8.8cm]{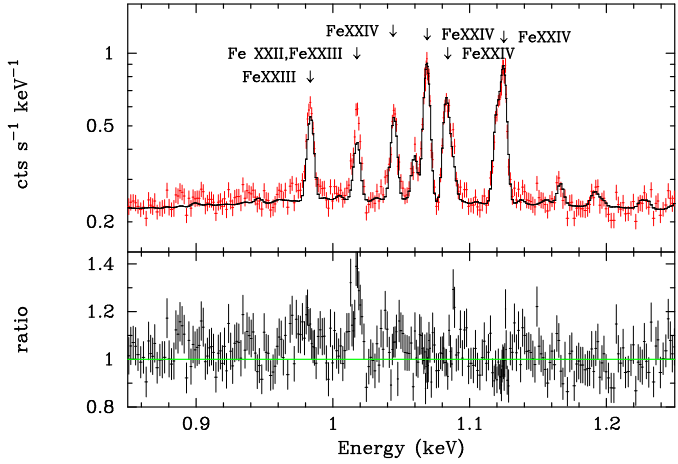}}
	\caption{Simulated XRISM spectrum based on best fitting model for A576, p=-100 ($\alpha=0.99$). The simulated exposure time is 500 ks. The spectrum is fitted with a one  temperature {\tt apec} model. In the top panel we show, in red, the simulated spectrum and in black the best fitting model. In the bottom panel  we show residuals in the form of a ratio of data to model. Note how the model reproduces adequately lines from Fe XXIV but only partially lines from XXII and XXIII. More specifically, only half of the equivalent width of  the line at 1.02 keV, corresponding to the 3p-$>$2s transition, is reproduced by the model. The spectrum  is shown in the observer frame.}
	\label{fig:sim1}
\end{figure}
In Fig.\ref{fig:sim1} we show a fit to the simulated data with a one temperature {\tt apec} model, the best fitting temperature is 3.7 keV. The figure zooms in on the 0.9-1.2 keV range where much of the line emission is concentrated.
\begin{table}
	\centering
	\caption{Major Low Temperature Emission Lines}
	\begin{tabular}{|c|c|c|c|}
		\hline
		\hline
		Ion      & Transition & Rest Frame En.$\mathrm{^{(a)}}$   & $kT$ of ME$\mathrm{^{(b)}}$  \\
		&            &  keV             &  keV   \\
		\hline
		Fe XXI      &  40 to 1  &  1.009  &  1.09       \\
		Fe XXIII    &  12 to 5  &  1.020  &  1.37       \\
		\hline
		Fe XXII     &  21 to 1  &  1.053  &  1.09       \\
		Fe XXIII    &  20 to 5  &  1.056  &  1.37       \\
		\hline
		Fe XXIV     &   4 to 3  &  1.085  &  1.72       \\
		Fe XXIV     &   8 to 3  &  1.109  &  1.72       \\
		\hline
		Fe XXIV     &   7 to 2  &  1.124  &  1.72       \\
		Fe XXIII    &  15 to 1  &  1.129  &  1.37       \\
		\hline
		Fe XXIV     &   5 to 1  &  1.163  &  1.72       \\
		Fe XXIV     &   6 to 1  &  1.168  &  1.72       \\
		\hline
		\hline
	\end{tabular}
	\begin{list}{}{}
		\item[Notes:]
		$\mathrm{^{(a)}}$ Here energies are expressed in the rest frame, in the text in the observer frame. 
		$\mathrm{^{(b)}}$ Temperature at which Maximum Emissivity (ME) is reached.
	\end{list}
	\label{tab:lines}
\end{table}
As we can see, the model can reproduce adequately several FeXXIV emission lines, however there are at least two  lines that are not entirely reproduced by the model. As indicated in the figure, see also Table \ref{tab:lines}\footnote{Information on all lines is drawn from the ATOMDB database which is also used to construct the {\tt apec} model.}, these are lines from Fe XXII and Fe XXIII which arise from gas at a temperature of $\sim$ 2 keV that is not present in the model. Fitting of the excess emission in the lines at 0.98 keV and 1.02 keV with Gaussians shows that in both instances they are detected at more than the 99.9\% confidence level. In the case of a shorter simulation of 100 ks the
excess emission is still detected, albeit at a somewhat smaller statistical significance, more specifically at the 99\% level for the 0.98 keV line and 99.9\% for the 1.02 keV line.

We have also fitted the data with the same model used as input for the simulation, the fit is a good one but we do not go into details here. Indeed the point we wish to convey with our simulation is not that we can recover the input model, but that the low ionization Fe emission lines associated to cool gas are detected.  

We have performed simulations for all objects in our sample, from best fits reported in Table \ref{tab:spec_ana}, for 
 $p=-100$ ($\alpha = 0.99$). 
In several instances we find that  the excess  in A576 can also be detected in other systems.  More specifically, in all clusters but one, A1650, the excess emission associated to the 3d-$>$2p transition at a rest frame energy of 1.053 keV (FeXXII) and 1.056 keV (FeXXIII), is detected at more than 99\% confidence, and, in all but two, at the 99.9\% confidence, see Table  \ref{tab:xifu}.

We point out that, given the relatively high surface brightness of the regions for which we simulate spectra, 
the background, which for the spectral range of interest is dominated by diffuse foreground emission from within our galaxy, does not play an important role in limiting our detections. The only instance where background could become an issue is for sources where one of the astrophysical  lines happens to fall within the same resolution element of one of the brightest foreground lines.
In summary, XRISM high resolution spectral observations have the power to detect and characterize emission from cool gas in all but one of our systems, if it is found in the quantities estimated from EPIC assuming the differential emission model described in Sect.\ref{sec:con_spec}.

\begin{figure}
	\centerline{\includegraphics[angle=-90,width=8.8cm]{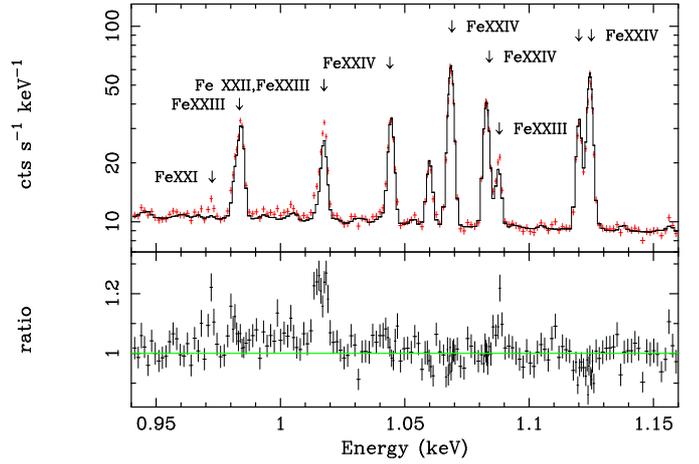}}
	\caption{Simulated XIFU spectrum based on best fitting model for A576, p=-100 ($\alpha=0.99$). The simulated exposure time is 50 ks. The spectrum is fit with a one  temperature {\tt apec} model. In the top panel we show, in red, the simulated spectrum and in black the best fitting model. In the bottom panel  we show residuals in the form of a ratio of data to model. Note how the model reproduces adequately lines from Fe XXIV but only partially lines from Fe XXII and XXIII. Interestingly, the combination of improved resolution and larger throughput with respect to XRISM, allows to detect excess emission at the Fe XXIII line at 1.09 keV and to detect an Fe XXI line
	at 0.97 keV. The spectrum  is shown in the observer frame.}
	\label{fig:sim2}
\end{figure}

\begin{table}
	\centering
	\caption{line detection significance  form XRISM (500 ks) and XIFU  (50 ks)}
	\resizebox{\columnwidth}{!}{%
		\begin{tabular}{|c| c |c c c|}	
			\hline
			\hline
			Name                    &   XRISM     & \multicolumn{3}{|c|}{XIFU}	  \\
			L.E.$^{\mathrm{^{a}}}$  & 	 1.053/1.056	  & 1.009 &	1.053/1.056	&	1.129	   \\
                                    & 	 \%	              &  \%   &	 \%	&	\%	   \\
			\hline
			A1644      &    $>99.9$	   &	$>99.9$ & $>99.9$ &  $>99.9$	\\
			A1650      &    -	       &	-       & - 	  & -     \\    
			A1689      &    $>99.9$	   &	$>99.9$ & $>99.9$ & $>99.9$  \\     
			A2256      &    $>99.9$	   &  $>99.9$   & $>99.9$ & $>99.9$   \\
			A3558      &   	$>99.9$	   &  $>99.9$   & $>99.9$ & $>99.9$	\\	
			A3562      &   $>99$	   &    - 	    & $>99.9$ & - 	\\
			A3571      &   $>99.9$	   &  $>95$     & $>99.9$ & $>99.9$ \\
			A3667      &   	$>99.9$	   &  $>99.9$   & $>99.9$ & $>99.9$ 	\\
			A4038      &    $>99.9$	   &  $>99.9$   & $>99.9$ & $>99.9$   \\
			A576       &   $>99.9$	   &  $>99.9$   & $>99.9$ & $>99.9$    \\
			A754       &   $>99.9$	   &  $>99.9$   & $>99.9$ & $>99.9$    \\
			MKW3s      &   $>99.9$	   &  $>99.9$   & $>99.9$ & $>99.9$   \\
			\hline
			\hline
		\end{tabular}%
	}
	\begin{list}{}{}
		\item[Notes:]
	$\mathrm{^{(a)}}$ line energy in keV in the source rest frame.
	\end{list}
	\label{tab:xifu}
\end{table}

In the case of the ATHENA XIFU experiment \citep{Barret_XIFU:2013} we have made use of spectral response and background files as made available at the web page\footnote{{\tt \small http://x-ifu.irap.omp.eu/resources/for-the-community}}.
As a first exercise, we have performed a 50 ks simulation on A576 based on the same model adopted for the XRISM case. As for XRISM, we fit the simulated data with a single temperature model and found at least three lines, the one previously identified from the XRISM simulations, and another two that are not fully accounted for by the model. 
In the case of the FeXXII and Fe XXIII line at 1.02 keV and the Fe XXIII line at 1.09 keV the single temperature model partially reproduced the line, while in the case of the Fe XXI line at 0.97 keV the line is entirely absent from the model. By modeling the excesses/lines with gaussians we find they are all detected at significance greater than 99.9\%, see Fig.\ref{fig:sim2} and Table \ref{tab:xifu}. Clearly, the much larger throughput of the ATHENA telescope, when compared to the XRISM one, allows for better statistics on the emission lines. As for XRISM, we have performed simulations of the best fitting spectral models of all objects in our sample, for $p=-100$ ($\alpha = 0.99$). In almost every case where the simulation includes emission from gas with temperatures  below $\simeq 2$ keV the Fe XXI, Fe XXII and Fe XXIII lines are detected at a very high significance,  see Table \ref{tab:xifu}.
Note that, in the case of  A1650, no line emission from cooler gas is detected either with XRISM or XIFU for the simple reason that the minimum temperature of the {\tt wdem} model on which the simulations are based is 3.45 keV, see Table \ref{tab:spec_ana}. In summary, XIFU high resolution spectral observations have the power to detect and characterize emission from cool gas in all  CCRs where such gas exists.

\section{Summary}\label{sec:summary}

In this paper we have investigated the relationship between CC and NCC systems, these are our main findings.

\begin{itemize}
	\item We have indirect evidence that NCCs can turn into CCs, the  most compelling is that the CC fraction is stable across cosmic time \citep{Ruppin:2021} despite the high rate of mergers disrupting CCs.
	\item By fitting a multi-temperature model to the central regions of CCRs we find that the majority of these systems are consistent with hosting small amounts of low temperature gas.
	\item Assuming pressure equilibrium between the different phases of the multi-temperature model we find that a sizable fraction of the gas can have entropy and cooling times that are substantially smaller than those found under the assumption of a single temperature.
	\item A comparison with entropy and cooling time distributions in cores of CC systems shows they are not too dissimilar from those found in the majority of our CCRs.
	\item If we allow CCRs to evolve adiabatically during the reconstitution of radial pressure gradients, the cooling time distributions become very similar to those observed in CC systems.   
	\item Considering that the number of CCs is stable across cosmic time and assuming, as suggested by our analysis, that the timescale over which a CC is reformed is comparable to the timescale over which mergers are active (e.g. duration of RHs), we should observe  a number of systems in the CC reconstitution phase comparable to the number of systems undergoing mergers. We suggest that some of our CCRs may actually be undergoing such a transformation.
    \item Within the evolutionary model we propose there is no net separation between CCs and NCCs. This naturally explains the presence of intermediate objects as well as conflicting results sometimes returned by different classification schemes.	
	\item High spectral resolution detectors to fly on board XRISM and ATHENA will have the power to resolve out the Fe L-shell emission lines coming from low temperature gas. This will allow us to firmly establish if some NCCs carry in their core short cooling time gas which permits the transition to CCs.   
\end{itemize}
We close by pointing out that CCs and NCCs should not be viewed as distinct sub classes, but as ``states" between which clusters can move.
\acknowledgements We acknowledge financial support from INAF mainstream project No.1.05.01.86.13. We acknowledge use of the ACCEPT Archive. MG acknowledges partial support by HST GO-15890.020/023-A and the {\it BlackHoleWeather} program. We thank an anonymous referee for valuable comments on a previous version of this manuscript.

\bibliography{biblio_icm_evol}

\appendix
\section{Deriving Thermodynamic properties for the isobaric model}
\label{sec:app}

In this Appendix we describe the  derivation of the gas mass versus entropy distribution presented in Eq.\ref{eq:m_vs_k}. 
We derive the thermodynamic properties of a multi-temperature plasma where the different phases are in pressure equilibrium and emit according to the relation: 

\begin{equation}
	dEM = EM_* \Big( {T\over T_{max}}\Big)^{\alpha-1} {dT \over T_{max}} \, ,
	\label{eq:dem_ap}
\end{equation}
where $dEM$ is the differential emission measure associated to the plasma at temperature $T$, 
$T_{max}$ is the maximum temperature, $\alpha$ parametrizes the slope of distribution and $EM_*$ its normalization.

We start by recalling the definition of the differential emission measure: 
\begin{equation}
	dEM = n_{H} n_{e} dV\, ,
	\label{eq:dem_def}
\end{equation}
where $n_{H}$ and $n_{e}$ are respectively the hydrogen and electron density and $dV$ is the differential volume.
We search for solutions of the form:

\begin{equation}
	V(<T) = V_{o}  \Bigl({T \over T_{max}}\Bigr)^\beta\, ,
	\label{eq:vo}
\end{equation}

where $V_{o}$ is the total volume of the region containing the multi-phase plasma, $T_{max}$ is the maximum temperature and  $	V(<T)$ is the volume occupied by all phases with temperature smaller than $T$. The parameters that need to be derived to characterize the distribution are the normalization $V_{o}$ and the slope $\beta$. 
By defining $t \equiv T/T_{max}$, differentiating  Eq.\ref{eq:vo} and substituting it into  Eq.\ref{eq:dem_def} 
we get 
\begin{equation}
	dEM = n_H n_e V_o dt^{\beta} \, .
	\label{eq:t1}
\end{equation}

By equating Eq.\ref{eq:t1} to Eq.\ref{eq:dem_ap} we get:

\begin{equation}
	n_H n_e V_o dt^{\beta} = EM_* t^{\alpha -1}dt\, .
	\label{eq:t2}
\end{equation}

By making use of the equations of state:  $p_H = n_H T $ and $p_e = n_e T $  we rewrite this as:

\begin{equation}
	n_{H,o} n_{e,o} t^{-2} V_o dt^{\beta} = EM_* t^{\alpha -1}dt\, ,
	\label{eq:t3}
\end{equation}

where $n_{H,o} = p_H/T_{max}$ and   $n_{e,o} = p_e/T_{max}$.
By carrying the calculation through, performing the derivation and solving for $V_o$ we get:

\begin{equation}
	  V_o  = {EM_* \over \beta \, n_{H,o} n_{e,o} }  t^{\alpha -\beta + 2}\, .
	\label{eq:t4}
\end{equation}

Since in our model we assume that $V_o$ does not depend on the temperature, we find the following solution:

\begin{align}	
	\beta &= \alpha +2  \, , 	\label{eq:t5} \\ 
	 V_o  &= { \alpha \over \alpha + 2} {EM \over n_{H,o} n_{e,o}}  \, , \label{eq:t6}
\end{align}
where we have substituted $EM_*$ with $ \alpha EM$, see Eq.\ref{eq:em_norm}. This finally leads to the expression of the  volume as a function of temperature:
  
  \begin{equation}
  	V(<T) =  V_o \Bigl({T \over T_{max}}\Bigr)^{\alpha+2}\, ,
  	\label{eq:vt}
  \end{equation}
where $V_o$ is given in Eq.\ref{eq:t6}.
Rearranging  Eq.\ref{eq:t6}, we can solve for $n_{e,o}$:
\begin{equation}
	n_{e,o}  =   r^{1/2} \,\Bigl({ \alpha \over \alpha + 2}\Bigr)^{1/2}\Bigl({ EM \over V_o  }\Bigr)^{1/2} \, ,
	\label{eq:t7}
\end{equation}
where  $r \equiv n_e / n_H $, for which we assume a standard value of $r=1.13$. Since our model is isobaric, once the density and temperature of the hottest phase are known, the density of any other can be computed from the relation:
\begin{equation}
	n_{e}  =   n_{e,o} \Bigl({ T \over T_{max}}\Bigr)^{-1} \, .
	\label{eq:t8}
\end{equation}

Having expressed density as a function of temperature we can easily do the same for entropy, with a little algebra we find:

\begin{equation}
 	K = K_{o}  \Bigl({T \over T_{max}}\Bigr)^{5/3}\, ,
 	\label{eq:k_t}
\end{equation}
where 
\begin{equation}
	K_{o} = T_{max}/n_{e,o}^{2/3} \, .
	\label{eq:k_o}	
\end{equation}

As already stated, we are interested in expressing the gas mass as a function of the entropy. 

We compute the volume as a function of the entropy, we proceed by inverting Eq.\ref{eq:k_t} and substituting into Eq.\ref{eq:vt},

\begin{equation}
	V(<K) =  V_o \Bigl({K \over K_{o}}\Bigr)^{{3\over 5}(\alpha+2)}\, .
	\label{eq:vk}
\end{equation}
From this we also compute the derivative of the volume with respect to the entropy:
\begin{equation}
	{dV\over dK} =  {3\over 5}(\alpha+2) \, {V_o \over K_o} \,  \Bigl({K \over K_{o}}\Bigr)^{{1\over 5}(3\alpha+1)}\, .
	\label{eq:dvdk}
\end{equation}
Now we can compute the gas mass:

\begin{equation}
	M_{gas}(<V) =  \mu_e m_p \int_0^V n_e dV\, .
	\label{eq:mgv}
\end{equation}
where $\mu_e=1.12 $ is the  mean molecular weight per free electron,  and $m_p$ the proton mass.
By replacing $dV$ with $dV/dK \cdot dK$, making use of Eqs.\ref{eq:t8}  and \ref{eq:dvdk} we get:
\begin{equation}
	M_{gas}(<K) =  {3\over 5}(\alpha+2) \,  \mu_e m_p n_{e,o} V_o \int_0^K   \Bigl({\mathcal{K}\over \mathcal{K}_o}\Bigr)^{{1\over 5}(3\alpha-2)} {d\mathcal{K} \over \mathcal{K}_o}\, ,
	\label{eq:mgk}
\end{equation}
where we used the symbol ${\mathcal{K}}$ for the entropy in the integral to avoid confusion with $K$, which denotes the upper limit of integration.
By carrying out the integration we find:

\begin{equation}
	M_{gas}(<K) = M_{gas,t}  \Bigl({K \over K_{o}}\Bigr)^{{3\over 5}(\alpha+1)}\, ,
	\label{eq:m_gas2}
\end{equation}
where
\begin{equation}
	M_{gas,t}= {\alpha + 2\over \alpha + 1} \,  \mu_e m_p n_{e,o} V_o \, .
	\label{eq:m_gas_tot}
\end{equation}
Eq.\ref{eq:m_gas2} applies to the case where the minimum temperature of the 
multiphase medium is 0, i.e. $T_{min} = 0 $. It may be easily generalized to the case of an arbitrary  $T_{min}$ simply by substituting 0 in the lower limit of integration in Eq.\ref{eq:mgk} with $K_{min}$, which is derived from Eq.\ref{eq:k_t} by setting $T=T_{min}$.
In this case the gas mass can be expressed as:
\begin{multline}
	M_{gas}(K>K_{min}) =  \\ M_{gas,t}
	\Biggl[ \Biggl( {K \over K_o}\Biggr)^{{3\over 5}(\alpha +1)} - \Biggl( {K_{min} \over K_o}\Biggr)^{{3\over 5}(\alpha +1)}\Biggr]  \, ,
	\label{eq:m_gas3}
\end{multline}

which is the same equation reported in Eq.\ref{eq:m_vs_k}.

The total volume of the emitting region for the case $T_{min} > 0 $,  can be derived from Eq.\ref{eq:vt} as the difference between the volume associated with phases with $T<T_{max}$ and the one associated  with phases with $T<T_{min}$,
\begin{equation}
	V_{tot} =  V_o \Biggl[1 - \Biggl({T_{min} \over T_{max}}\Biggr)^{\alpha+2}\Biggr]\, .
	\label{eq:vtot}
\end{equation}

Similarly, the emission measure for the case $T_{min} > 0 $,  can be derived by 
integrating the differential emission measure, see Eq.\ref{eq:dem_ap} from $T_{min}$ to $T_{max}$:
\begin{equation}
	EM(T_{min},T_{max}) = EM  \, \alpha \int_{T_{min}}^{T_{max}}  \Big( {T\over T_{max}}\Big)^{\alpha-1} {dT \over T_{max}} \, ,
	\label{eq:em_ap}
\end{equation}
where we have substituted $EM_*$ with $EM \, \alpha$, see Eq.\ref{eq:em_norm}. By integrating Eq.\ref{eq:em_ap} we get:
\begin{equation}
	EM(T_{min},T_{max}) = EM  \,  \Biggl[ 1 - \Biggl({T_{min}\over T_{max}}\Biggr)^{\alpha}  \Biggr]\, .
	\label{eq:em2_ap}
\end{equation}

From a practical point of view, this is how we go from results of the spectral fits to physical quantities:
using Eq.\ref{eq:em2_ap} we take measured values of $EM(T_{min},T_{max})$, $\alpha$, $T_{min}$ and $T_{max}$ and compute $EM$. Using Eq.\ref{eq:vtot} we go from $V_{tot}$, which we estimate from the angular size of the region from which the spectrum has been extracted\footnote{To estimate a volume we also need a value for the extension in the radial direction, in the absence of any measure, we assume this to be  comparable to that in the plane of the sky, in practice $\Delta r = \sqrt A$, where $A$ is the area in the plane of the sky and $\Delta r$ is the extension of the volume in the radial direction}, to $V_o$. Using Eq.\ref{eq:t7} we compute $n_{e,o}$ from   $\alpha$, $V_o$ and $EM$ and using Eq.\ref{eq:k_o} we compute $K_o$ from $n_{e,o}$ and $T_{max}$. From these quantities all others can be easily derived.

Finally, to compute the gas mass vs. cooling time distribution we adopt the standard isobaric cooling time formula which includes adiabatic compression, hence the 5/2 factor instead of 3/2: 
 \begin{equation}
 	t_{cool} = { 5 \over 2} {n_{gas} kT \over n_e n_H \Lambda(T,Z)} \, ,
 	\label{eq:t_cool}
 \end{equation}
 where $n_{gas}$ is the gas density and $\Lambda(T,Z)$ is the cooling function. The gas mass vs. cooling time distribution is computed numerically, for any given value of $M_{gas}$ we use the set of equations presented above to derive the associated values of $T$, $n_{gas}$ and $n_{e}$, the value of $Z$ is constant and comes from the spectral fit of the source under consideration. We plug the above values into Eq.\ref{eq:t_cool} and derive the associated cooling time.

\section{Deriving the gas mass versus cooling time in 
	the CC reformation phase}
\label{sec:appb}

We wish to derive a gas mass vs. cooling time relation that is consistent with the gas mass vs. entropy relation 
derived in the framework of the isobaric model and with radial profiles of the form:

\begin{equation}
	p = p_o \Bigl({r \over r_{max}}\Bigr)^\epsilon \,\,\, {\rm and} \,\,\, K = K_o \Bigl({r \over r_{max}}\Bigr)^\gamma \, ,  
	\label{eq:p_k_ap2}
\end{equation}

where $p_o = n_o T_{max}$ and $r_{max}$ is estimated from the size of the region from which spectra are extracted.
From the entropy profile in Eq.\ref{eq:p_k_ap2} and Eq.\ref{eq:m_gas2} we derive the gas mass radial profile:

\begin{equation}
	M_{gas}(<r) = M_{gas,t}  \Bigl({r \over r_{max}}\Bigr)^{{3\over 5} \gamma (\alpha+1)}\, .
	\label{eq:m_gasr}
\end{equation}

As for Eq.\ref{eq:m_gas2} this equation can be easily generalized to the $T_{min} \ne 0 $ case:

\begin{multline}
	M_{gas}(r>r_{min}) = \\  \, M_{gas,t}  
	\Biggl[ 
	\Bigl({r \over r_{max}}\Bigr)^{{3\over 5} \gamma (\alpha+1)} -  
	\Bigl({r_{min} \over r_{max}}\Bigr)^{{3\over 5} \gamma (\alpha+1)}\Biggr] \, ,
	\label{eq:m_gasr2}
\end{multline}
where $r_{min}$ is derived from Eq.\ref{eq:p_k_ap2} by imposing $K=K_{min}$.

From the pressure and entropy radial profiles in Eq.\ref{eq:p_k_ap2} we can use standard thermodynamic formulae to 
recover radial profiles for density and temperature; by inverting these profiles and substituting into Eq.\ref{eq:m_gasr} we can derive the relation between the gas mass and any thermodynamic variable. 
For instance, the gas mass vs. temperature relation is given by:

\begin{equation}
	M_{gas}(<T) = M_{gas,t}  \Bigl({T \over T_{max}}\Bigr)^{{3\gamma \over 2\epsilon + 3\gamma} (\alpha+1)}\, ,
	\label{eq:m_gast}
\end{equation}

and the gas mass vs. Hydrogen density relation is given by:

\begin{equation}
	M_{gas}(<n_{H}) = M_{gas,t}   \Bigl({n_{H} \over n_{H,o}}\Bigr)^{{\gamma \over \epsilon - \gamma} (\alpha+1)} \, .
	\label{eq:m_gasnh}
\end{equation}

Both Eq.\ref{eq:m_gast} and \ref{eq:m_gasnh} can be generalized to the $T_{min} \ne 0 $ case:

\begin{multline}
	M_{gas}(T>T_{min}) = \\ M_{gas,t} 
	\Biggl[ 
	\Bigl({T \over T_{max}}\Bigr)^{{3\gamma \over 2\epsilon + 3\gamma} (\alpha+1)} - 
	\Bigl({T_{min} \over T_{max}}\Bigr)^{{3\gamma \over 2\epsilon + 3\gamma} (\alpha+1)}
	\Biggr] 
	\, ,
	\label{eq:m_gast2}
\end{multline}

\begin{multline}
	M_{gas}(n_{H}>n_{H,min}) = \\ M_{gas,t}  
	\Biggl[ 
	\Bigl({n_{H} \over n_{H,o}}\Bigr)^{{\gamma \over \epsilon - \gamma} (\alpha+1)} -
	\Bigl({n_{H,min} \over n_{H,o}}\Bigr)^{{\gamma \over \epsilon - \gamma} (\alpha+1)} 
	\Biggr] 	 
	\, .
	\label{eq:m_gasnh2}
\end{multline}

From Eqs.\ref{eq:m_gast2} and \ref{eq:m_gasnh2} using the expression for the cooling time reported above (see Eq.\ref{eq:t_cool}) we can proceed to compute numerically the gas mass versus cooling time relation.

\end{document}